\documentclass[prd,twocolumn,showpacs,nofootinbib,preprintnumbers,floatfix]{revtex4} 
\usepackage[utf8]{inputenc}
\usepackage{amsfonts,amsmath,amssymb}
\usepackage{graphicx}
\usepackage{color}
\usepackage{lipsum}

\usepackage{caption}
\usepackage[plainpages=false, colorlinks=true, anchorcolor=blue, linkcolor=blue, citecolor=blue, bookmarks=false]{hyperref}
\usepackage{natbib}
\usepackage{enumitem}
\usepackage{subcaption}
\newcommand{\rthis}[1]{\textcolor{black}{#1}}
\begin{document}
\newcommand{\apjl}{Astrophys. J. Lett.}
\newcommand{\apjs}{Astrophys. J. Suppl. Ser.}
\newcommand{\aap}{Astron. \& Astrophys.}
\newcommand{\aj}{Astron. J.}
\newcommand{\araa}{Ann. Rev. Astron. Astrophys. } %ARA$\&$A}
\newcommand{\aapr}{Astronomy and Astrophysics Review}
\newcommand{\mnras}{Mon. Not. R. Astron. Soc.}
\newcommand{\apss} {Astrophys. and Space Science}
\newcommand{\jcap}{JCAP}
\newcommand{\pasj}{PASJ}
\newcommand{\pasa}{Pub. Astro. Soc. Aust.}
\newcommand{\physrep}{Physics Reports}
\newcommand{\ssr}{Space Science Reviews}

\title{Constraints on Self-Interacting dark matter from relaxed galaxy groups}
\author{Gopika \surname{K.}}\altaffiliation{E-mail:ph19resch01001@iith.ac.in}

\author{Shantanu  \surname{Desai}}  
\altaffiliation{E-mail: shntn05@gmail.com}

\begin{abstract}
Self-interacting dark matter (SIDM) has been proposed as an alternative to the standard collisionless cold dark matter to explain the diversity of galactic rotation curves and core-cusp problems seen at small scales. Here, we estimate the constraints on SIDM for a sample of 11 relaxed galaxy groups with X-ray observations from  Chandra and XMM-Newton. We fit the dark matter density distribution to  the Einasto profile and use the estimated Einasto $\alpha$ parameter to constrain the SIDM cross-section, based on the empirical relation between the two, which was obtained in ~\cite{Eckert22}.  We obtain  a  non-zero central estimate for the cross-section per unit mass ($\sigma/m$) for seven groups, with   the most precise estimate obtained for NGC 5044, given by $\sigma/m=0.165 \pm 0.025~\rm{cm^2/g}$, for dark matter velocity dispersion of about 300 km/sec. For the remaining four groups, we  obtain 95\% c.l. upper limits on   $\sigma/m  < 0.16-6.61~\rm{cm^2/g}$  with dark matter velocity dispersions between  200-500 km/sec, with the most stringent limit for our sample obtained for the  group  MKW 4,  given by $\sigma/m< 0.16~\rm{cm^2/g}$ for dark matter velocity dispersion of about 350 km/sec.
\end{abstract}

\affiliation{Department of Physics, Indian Institute of Technology, Hyderabad, Telangana-502284, India}
\maketitle

\section{Introduction}
The dark matter problem is one of the most vexing problems in modern physics and astrophysics. Although dark matter constitutes about 25\% of total energy density of the universe and is a basic tenet of the $\Lambda$CDM model~\cite{Planck20}, its identity is still unknown, despite close to 100 years of evidence~\cite{Hooper}. There is also no laboratory evidence for some of the most well motivated dark matter candidates such as Weakly Interacting Massive Particles (WIMPs) or axions or through indirect searches, which have only reported null results~\cite{Merritt,Gaskins,Desai04,Boveia,Schumann,HESS,MAGIC,ChanLee}.
Furthermore, the currently established $\Lambda$CDM model has also faced some tensions at scales smaller than 1 Mpc, such as the core/cusp problem~\cite{Primack94,Moore94,DelPopolo22}, missing satellite problem~\cite{Bullock,Weinberg}, too big to fail problem~\cite{Boylan11}, and  satellites plane problem~\cite{Banik}. Data on galactic scales  for spiral galaxies have also revealed some intriguing deterministic scaling relations or correlations  such as the Radial Acceleration Relation (RAR) ~\cite{McGaugh16,our_rar_2020}, mass acceleration discrepancy relation~\cite{LRR},  the constancy of dark matter halo surface density~\cite{Donato09}, linearity of dark matter to baryonic matter ratio~\cite{Lovas}, which remain an enigma, although recent works have  shown that some of these  regularities could  be explained within $\Lambda$CDM~\cite{ps21,Gopika23,Sales23},  while  some other observations such as the  RAR~\cite{Chan20,Tian,our_rar_2020,Pradyumna21,ChanDesai}, constancy of dark matter surface density~\cite{Gopika20,Gopika2021}  and linearity of dark to baryonic matter ratio~\cite{Varenya} are  not universal. Therefore a large number of alternatives to the standard $\Lambda$CDM model have emerged such as Self-Interacting Dark Matter (SIDM, hereafter)~\cite{SIDM}, Superfluid dark matter~\cite{Khoury}, Warm Dark Matter~\cite{Colin}, Wave (or Fuzzy) Dark Matter~\cite{EFerreira}, Flavor Mixed Dark Matter~\cite{Medvedev}, modified gravity (which obviates the need for dark matter)~\cite{LRR,Banik}, etc.

The original motivation of SIDM about twenty years ago  was to resolve the core/cusp and missing satellite problems~\cite{Spergel2000}. Although the missing satellite problem is no longer a major point of contention given the discovery of many satellite galaxies of the Milky way from wide-field sky surveys~\cite{Nadler,SIDM}, the core/cusp problem is still an issue~\cite{Oh11}. Furthermore, another acute problem is the diversity in the  dark matter density profiles inferred from rotation curves, which cannot be resolved using feedback models~\cite{Kaplinghat20}.  SIDM can resolve these problems and also some of the aforementioned anomalies  at the galactic scale such as RAR and constant dark matter surface density~\cite{Tulin,Kaplinghat,SIDM,Ren19}.
SIDM has been applied to a whole suite of astrophysical observations from dwarf galaxies to galaxy clusters (See ~\cite{SIDM,Tulin} for recent reviews). Current observations are consistent with a velocity-dependent SIDM cross-sections with  values of $\sigma/m \sim 2~\rm{cm^2/g}$ on single galaxy scales to    $\sigma/m \sim 0.1~\rm{cm^2/g}$ on galaxy cluster scales~\cite{Kaplinghat15}.

A large number of works have obtained results on SIDM cross-sections using galaxy clusters~\cite{Meneghetti01,Gnedin01,Mahdavi07,Randall08,Bradac08,Peter13,Kahl14,Gastaldello14,Massey15,Harvey15,Robertsonbullet,Harvey19,Banerjee20,Sagunski,Eckert22} as well as a combination of isolated galaxies and brightest cluster galaxies (BCGs) within clusters~\cite{Yang23}.  These include looking at the  shape distribution of galaxy cluster arcs~\cite{Meneghetti01}, dark matter-galaxy offsets~\cite{Randall08,Robertsonbullet}, spherical Jeans modeling~\cite{Sagunski}, subhalo evaporation of elliptical galaxies in clusters~\cite{Gnedin01}, fitting the dark profiles  to cored isothermal~\cite{Andrade} or Einasto profiles~\cite{Eckert22}, offset between dark matter and stars~\cite{Massey15,Harvey15,Kahl14}, baryon-dark matter separation~\cite{Gastaldello14}, offset between X-ray gas and galaxies~\cite{Bradac08}, offset between X-ray center and optical center~\cite{Adhikari23}, ellipticities of clusters~\cite{Peter13}, wobbling of BCG around the center of dark matter~\cite{Harvey19},  effect on the profiles at the outskirts near the splashback radius~\cite{Banerjee20}. The most stringent constraints come from the core densities obtained using cluster strong lensing with limits ranging from $\sigma/m <~\rm{0.13 ~cm^2/g}$~\cite{Andrade21} to $\sigma/m <~\rm{0.35~cm^2/g}$~\cite{Sagunski}. We also note that some of the earliest stringent limits obtained on the SIDM cross-section, for example  $\sigma/m<0.02~\rm{cm^2/g}$ from MS 2137-23~\cite{Miralda} had to be revised with the advent of SIDM simulations,  and subsequently  became much more relaxed with  $\sigma/m < 1~\rm{cm^2/g}$~\cite{Peter13}. Evidence for non-zero cross-sections have been  reported at cluster scales  with $\sigma/m \approx 0.19 \pm 0.09~\rm{cm^2/g}$~\cite{Sagunski}.
Most recently, an analysis 
of the offsets between the X-ray and central galaxy position for 23 clusters from DES and SDSS  is tentatively consistent with $\sigma/m \sim 1~\rm{cm^2/g}$~\cite{Adhikari23}. Similarly,   evidence for a non-zero cross section from galaxy groups has also been found  with    $\sigma/m = 0.5 \pm 0.2~\rm{cm^2/g}$~\cite{Sagunski}. 

In this work we obtain constraints on the SIDM cross-section  using a sample of  relaxed galaxy groups and low mass galaxy clusters imaged with Chandra and XMM-Newton~\cite{Gastaldello_2007}. Galaxy groups are dark matter dominated systems with masses in the range of $10^{13}-10^{14} M_{\odot}$ and containing less than 50 galaxies~\cite{Sun12,Lovisari,Oppenheimer}. The exact demarcation between a galaxy cluster and galaxy group is not well defined~\cite{Lovisari,Oppenheimer}.  On group scales the expected values of $\sigma/m$ are expected to be  $0.1-1~ \rm{cm^2/g}$~\cite{Kaplinghat15,Sagunski}. We have previously used these same systems to test the invariance of dark matter halo surface density and to test the RAR~\cite{Gopika2021}. To obtain constraints on SIDM cross-sections,  we follow the same methodology as ~\cite{Eckert22} (E22 hereafter). 

The outline of this manuscript is as follows. We recap the E22 analysis in Sect.~\ref{sec:E22}. The data sample used for this work along with  analysis procedure and results  are outlined in Sect.~\ref{sec:analysis}.  We conclude in Sect.~\ref{sec:conclusions}.

\section{E22 analysis}
\label{sec:E22}
Here, we provide an abridged summary of the method used in E22 to obtain limits on SIDM cross-section using galaxy clusters. For their analysis, they considered mock clusters with $M_{200}> 3 \times 10^{14} M_{\odot}$ from the {\tt Bahamas-SIDM} hydrodynamical cosmological simulations describing cluster formation~\cite{Bahamas,Robertson19}. These simulations have been carried out for four distinct values of $\sigma/m$ ranging from 0 (corresponding to CDM) to $1.0~\rm{cm^2/g}$ with a total of  230 clusters. These simulations also include baryonic feedback effects such as cooling, star formation, stellar as well as AGN feedback~\cite{Bahamas}. However, since there is no ab-initio understanding of how the first black holes form, for these simulations black hole seeds are injected by hand  into the dark matter halos, and the growth of black holes is governed by Bondi accretion following the prescription in ~\cite{Booth09}. These simulations do not model the cold interstellar medium, which could underestimate the accretion rate onto black holes. These simulations have been shown to reproduce the galactic stellar mass fraction including dependence on galaxy type, while also matching the hot gas fraction in groups and clusters~\cite{Bahamas}.  However, in these simulations, the stellar density profiles are sensitive  to the resolution and details of the AGN feedback processes~\cite{Robertson19}. A full listing of all the successes and shortcomings of these simulations have been  summarized in ~\cite{Bahamas,Robertson19}.
The synthetic  clusters from these simulations were then fitted to an Einasto profile~\cite{Einasto65}:
\begin{equation}
    \rho_{Ein}(r) = \rho_s \exp \bigg[-\frac{2}{\alpha} \bigg( \bigg( \frac{r}{r_s} \bigg)^\alpha  - 1 \bigg) \bigg]
    \label{eq:einasto}
\end{equation}
where $r$ is the radial distance; $r_s$ is the scale scale radius~\cite{Ettori19}; $\rho_s$ is the  scale density corresponding to the mass density at $r=r_s$; and  $\alpha$ is the Einasto index. The Einasto profile provided  a  good fit to the synthetic clusters.  
Based on these fits E22 constructed  an empirical relation between $\alpha$ and $\sigma/m$ which can be written as follows:
\begin{equation}
\alpha= \alpha_0 + \alpha_1 \left(\frac{\sigma/m}{1 cm^2/g}\right)^{\gamma}
\label{eq:alphasidm}
\end{equation}
where $\alpha_0$ is the mean value obtained for CDM, while $\alpha_1$ and $\gamma$ denote the scaling parameters that encode the dependence  of $\alpha$ on the SIDM cross section/unit mass. For relaxed clusters in hydrostatic equilibrium, E22 obtained $\alpha_0=0.178$, $\alpha_1=0.20$, and $\gamma=0.63$. This relation was found to be robust with respect to the choice of subgrid physics, which incorporates different  models of cooling, star formation, AGN and supernova feedback~\cite{Bahamas}. This empirical relation was then applied to the clusters in the  XMM-Newton Cluster
Outskirts Project (X-COP) sample~\cite{XCOP}, and a combined best fit value of $\alpha$ was used to constrain $\sigma/m$. The estimated value of $\sigma/m$ from this stacking analysis was found by E22 to be $\sigma/M < 0.19~\rm{cm^2/g}$ at 95\% c.l~\cite{Eckert22} at an assumed dark matter collision velocity of 1000 km/sec.
We should also point out that the isothermal Jeans profile~\cite{Kaplinghat15} could also adequately model the simulated SIDM halos~\cite{Robertson21}. However, in order to obtain constraints on SIDM cross-sections from isothermal profiles, one needs to make additional assumptions such as the age of the halo, etc~\cite{Robertson21}.

\section{Analysis and Results}
\label{sec:analysis}
We use X-ray observations of 17 relaxed galaxy groups imaged with Chandra and/or XMM-Newton with redshifts up to z = 0.08. The observational details for A2589 can be found in~\cite{Zappacosta_2006}, while details for all the remaining groups can be found in  \cite{Gastaldello_2007}. These groups have masses between ($10^{13}-10^{14}M_{\odot}$) and span the range between galaxies and clusters, with temperatures in the range 1-3 keV. These suites of galaxy groups were used to test the constancy of dark matter density and radial acceleration relation for groups in ~\cite{Gopika2021} as well as  the MOND paradigm~\cite{Angus}.

\subsection{Einasto fits to DM halos}
The reconstruction of the dark matter density profiles for the group sample considered in this work can be found in ~\cite{Gopika2021}. We only provide a brief summary here. We first estimate the total group mass after assuming  that the groups are in hydrostatic equilibrium, since we are dealing with relaxed systems. The gas mass was estimated  from the X-ray temperature and surface-brightness data. We then obtain the dark matter mass by  subtracting  the gas mass (using the fitted gas density profiles) and the stellar mass obtained from  the $K$-band luminosity for the brightest group galaxy. The dark matter density profile was then estimated from the dark matter mass  assuming spherical symmetry. The errors in the density profile have been obtained using error propagation based on the errors in the temperature. This error budget does not include systematic errors due to hydrostatic equilibrium and hydrostatic bias, spherical symmetry as well as due to uncertainties in the  stellar mass. These hydrostatic bias values range from no bias~\cite{Liu23} to about  40\%~\cite{Salvati} (see Ref.~\cite{Wicker} for  a  recent compilation of hydrostatic bias measurements in literature.) The latest state of the art cosmological simulations  predict a bias of about 20\%~\cite{Gianfagna}. In order to estimate the hydrostatic bias on a group by group basis, we would need robust lensing masses for all our objects, which are not available at the moment. E22 has shown that that the Einsato $\alpha$ parameter gets overestimated by 2-8\% based on the hydrostatic equilibrium assumption, depending on the model for non-thermal pressure support. The dynamical modeling in ~\cite{Gastaldello_2007} as well as in ~\cite{Angus} have been done assuming spherical symmetry. A detailed discussion of the validity of the spherical symmetrical assumption is discussed in one of our previous works and could cause systematic errors of about  5\% in the total mass determination~\cite{Gopika20}. It is also not straightforward to estimate this error for every group, since the X-ray images are intrinsically two dimensional in nature.  Finally, since the stellar mass  is sub-dominant compared to the gas and dark matter contributions, we neglect the uncertainties in the estimates of stellar mass. Other causes of systematic errors could be due to inadequate background modelling and subtraction and whose magnitude could be  about 5\%~\cite{Gastaldello_2007}.

Similar to E22, we fit  these density profile data for the groups to  the Einasto profile (Eq.~\ref{eq:einasto})  using three  free parameters: $\rho_s$, $r_{s}$, and $\alpha$. We  maximize the log-likelihood function using the {\tt emcee} MCMC sampler~\cite{emcee}.  The likelihood used for getting the best-fit parameters can be found in Eq.~\ref{eq:maxlike} in the Appendix. The priors for $\rho_{s}$ are in the range $10^{11} < \rho_{s} < 10^{16} M_{\odot}/\rm{Mpc^{-3}}$, $r_s$  between 0 kpc and 300 kpc, and $\alpha$ spans from 0 to 0.5.  
For the MCMC runs, the number of walkers was set to 200 and the total number of iterations to 5000, which attained a mean acceptance fraction of approximately 0.52 for these fits, where the acceptance fraction is defined as the fraction of the total  parameter space, which gets accepted for sampling the posterior pdf. The corner plots showing the 68\%, 90\%, and 99\% credible intervals for the best-fit parameters associated with the 11 galaxy groups used in this analysis can be found in the Appendix. We have excluded the groups ESO 3060170, MS 0116.3-0115, NGC 1550, NGC 4325, NGC 533, and RX J1159.8+5531 in this analysis as their density profiles could not be fitted  by  marginalized closed
contours for all the three  Einasto parameters. For these groups, at least one of the parameters  showed only  one-sided marginalized contours at 68\% credible intervals.
The dark matter density profiles  for the remaining    galaxy groups  along with the best-fit Einasto parameters is shown  in Figure~\ref{fig:Fig1}. For every group, we also show the normalized residuals in the bottom panel, where the residuals are normalized by the error in each data point. We find that two groups have a reduced $\chi^2> 2$.
Table~\ref{tab:Einasto} summarizes the parameters obtained from fitting the dark matter density profiles of the galaxy groups with an Einasto model along with their reduced $\chi^2$, which gives a measure of the efficacy of the fit.  
A graphical summary  of  the values of $\alpha$ for every group can be found in Fig.~\ref{fig:alpha}. 
The estimated values of $\alpha$ for all the systems are in the range of approximately 0.12-0.49.  This agrees with the  values obtained in E22 for the X-COP sample, which had found $\alpha \sim 0.19$. 

\begin{figure*}
     \centering
     \begin{subfigure}[b]{0.45\textwidth}
         \centering
         \includegraphics[width=\textwidth]{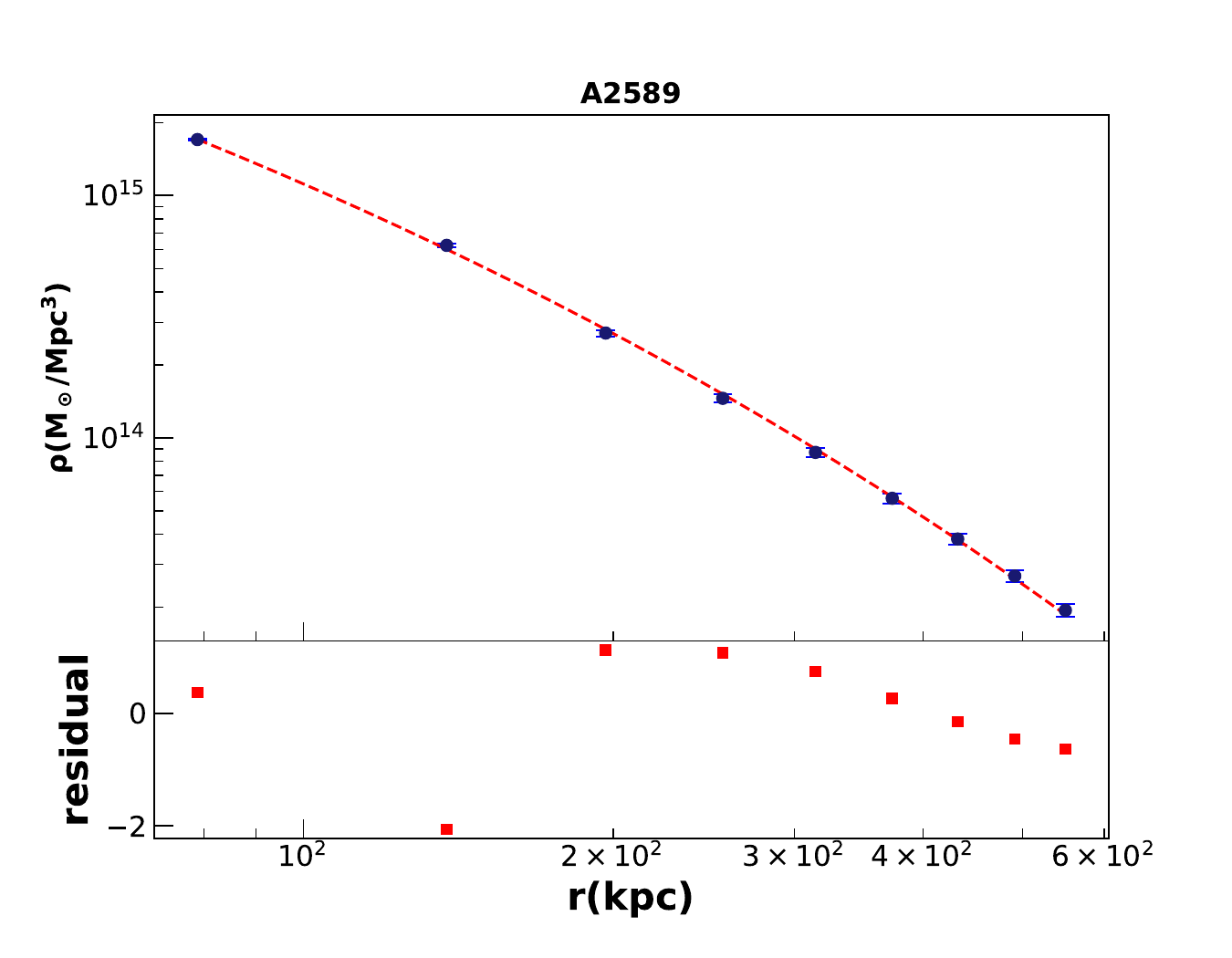}
         \label{figein:fig1.1}

     \end{subfigure}
     \hfill
     \hfill
     \begin{subfigure}[b]{0.45\textwidth}
         \centering
         \includegraphics[width=\textwidth]{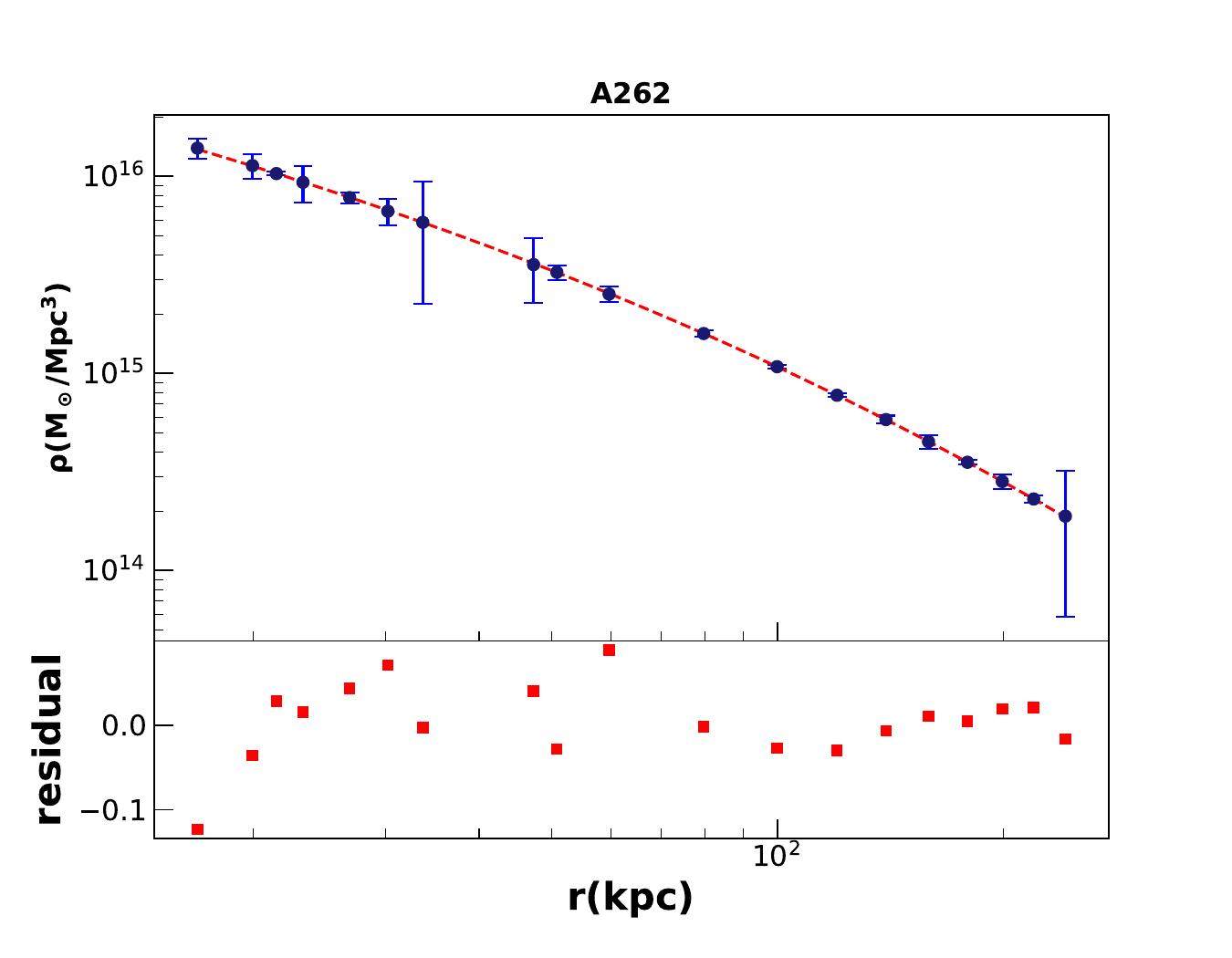}
         \label{figein:fig1.2}
     \end{subfigure}
        \hfill
          \hfill

     \begin{subfigure}[b]{0.45\textwidth}
         \centering
         \includegraphics[width=\textwidth]{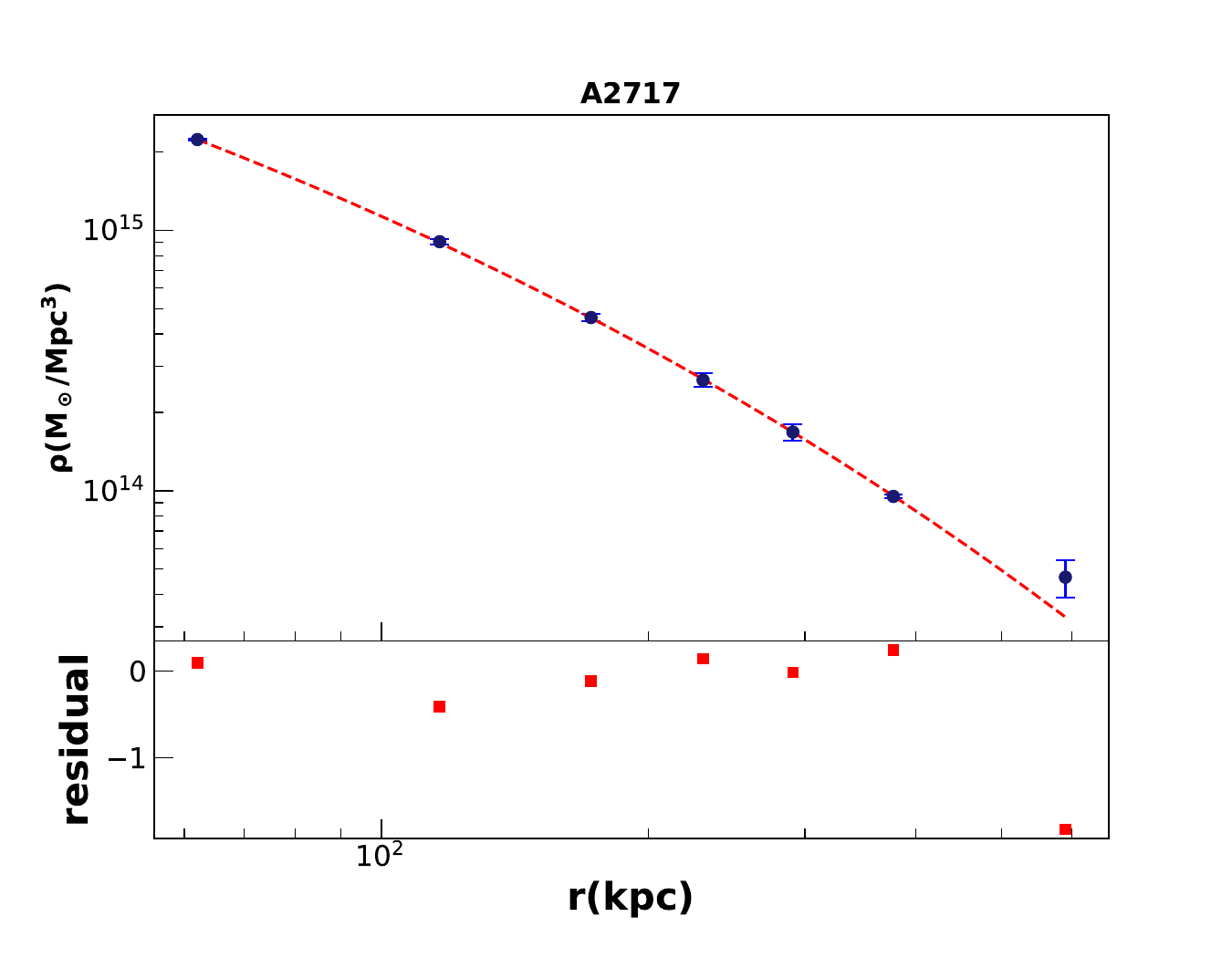}
         \label{figein:fig1.3}
     \end{subfigure}
     \hfill
     \hfill
     \begin{subfigure}[b]{0.45\textwidth}
         \centering
         \includegraphics[width=\textwidth]{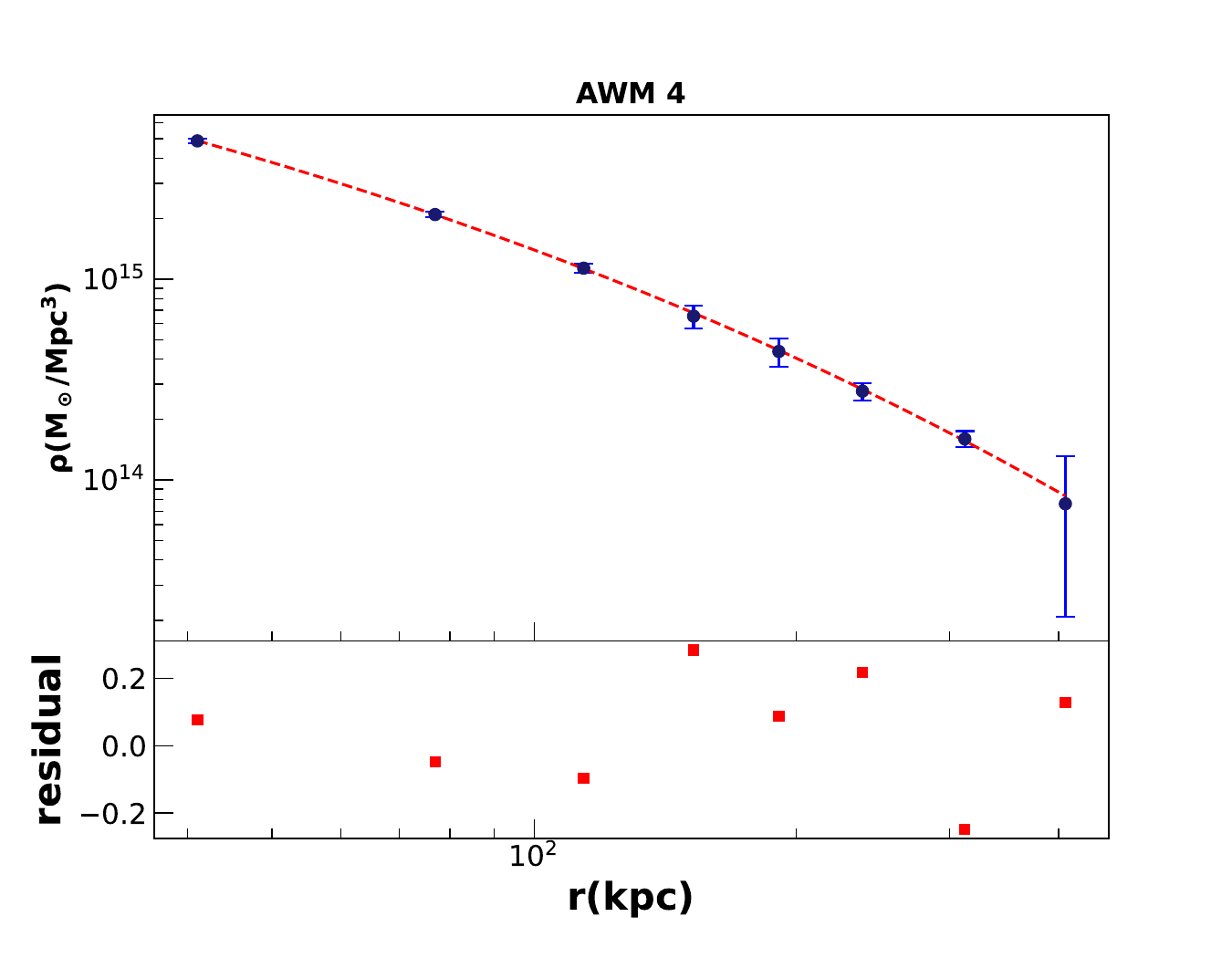}
         \label{figein:fig1.4}
     \end{subfigure}
     \hfill
     \hfill
    
      \begin{subfigure}[b]{0.45\textwidth}
         \centering
         \includegraphics[width=\textwidth]{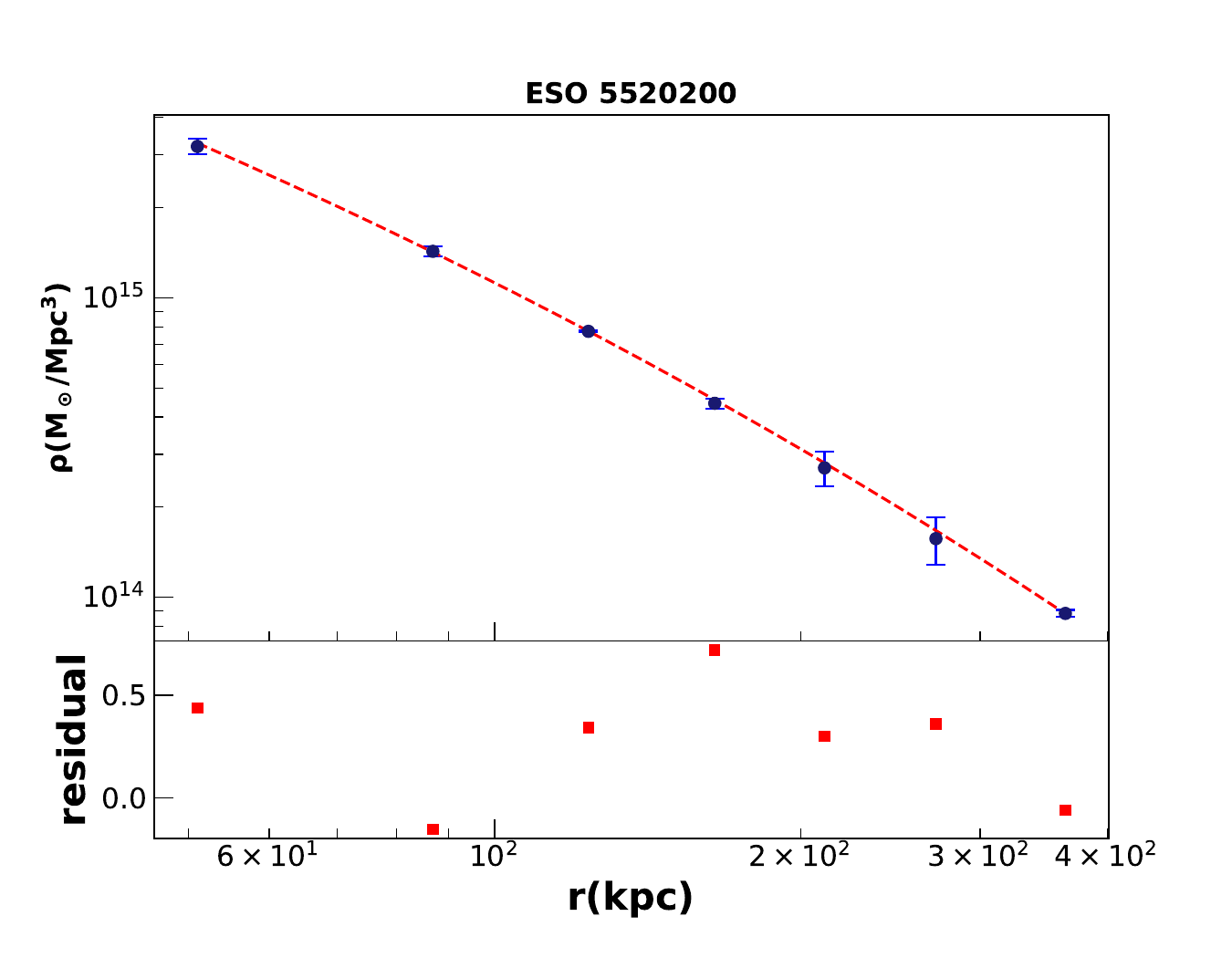}
         \label{figein:fig1.5}
     \end{subfigure}    
         \hfill
     \hfill   
           \begin{subfigure}[b]{0.45\textwidth}
         \centering
         \includegraphics[width=\textwidth]{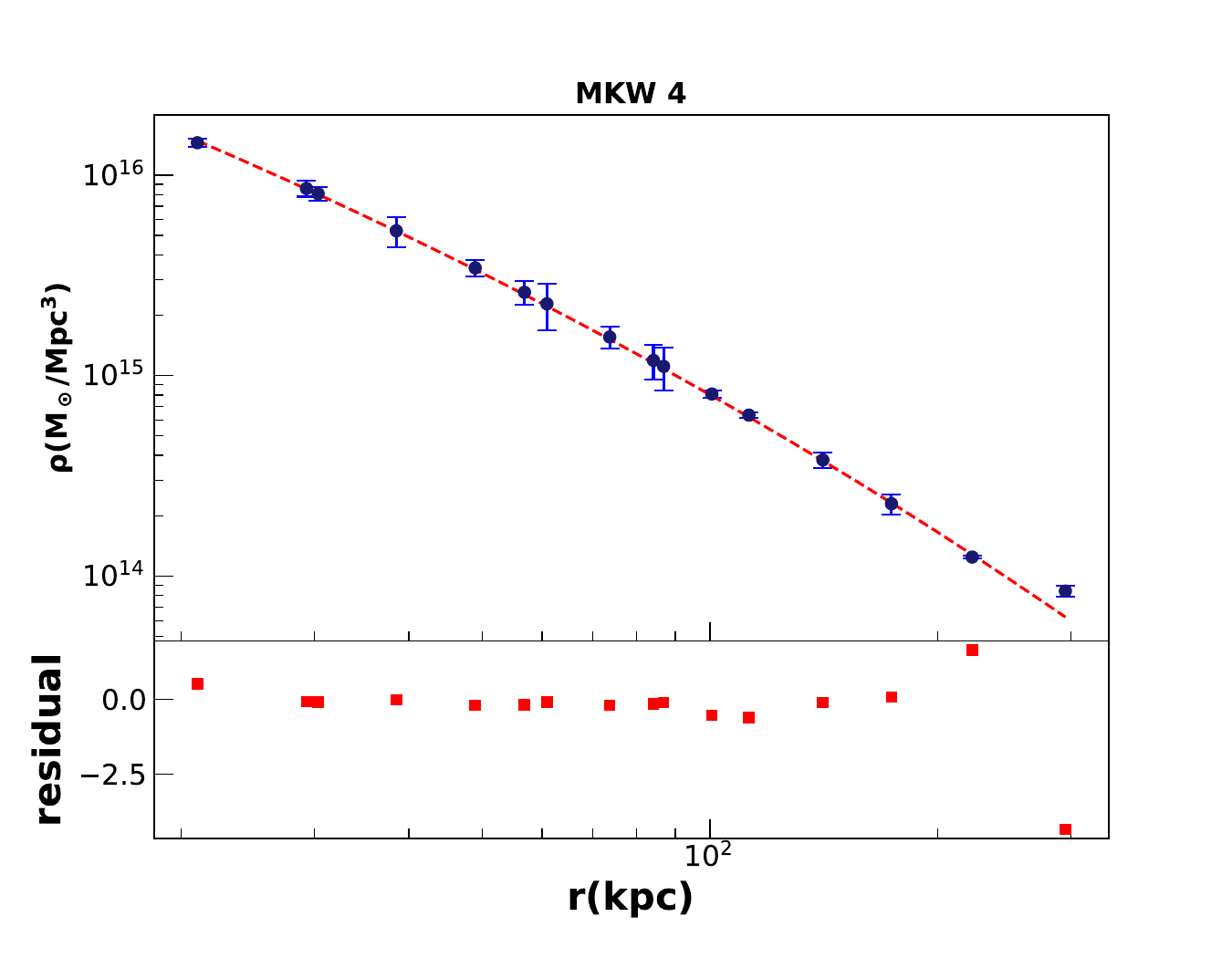}
         \label{figein:fig1.6}
     \end{subfigure}    
         \hfill
     \hfill
            \caption{Dark matter density profile for the galaxy groups fitted with an Einasto model. The top panel shows the dark matter density profile $\rho$ as a function of radius ($r$) from the group center along with the best-fit Einasto model parameters given in Table~\ref{tab:Einasto}. The bottom panel in each plot shows  the normalized residual given by the difference in the data and model divided by the error in the data.}

       \label{fig:Fig1}
\end{figure*}

\begin{figure*}\ContinuedFloat
     \centering
     
                \begin{subfigure}[b]{0.45\textwidth}
         \centering
         \includegraphics[width=\textwidth]{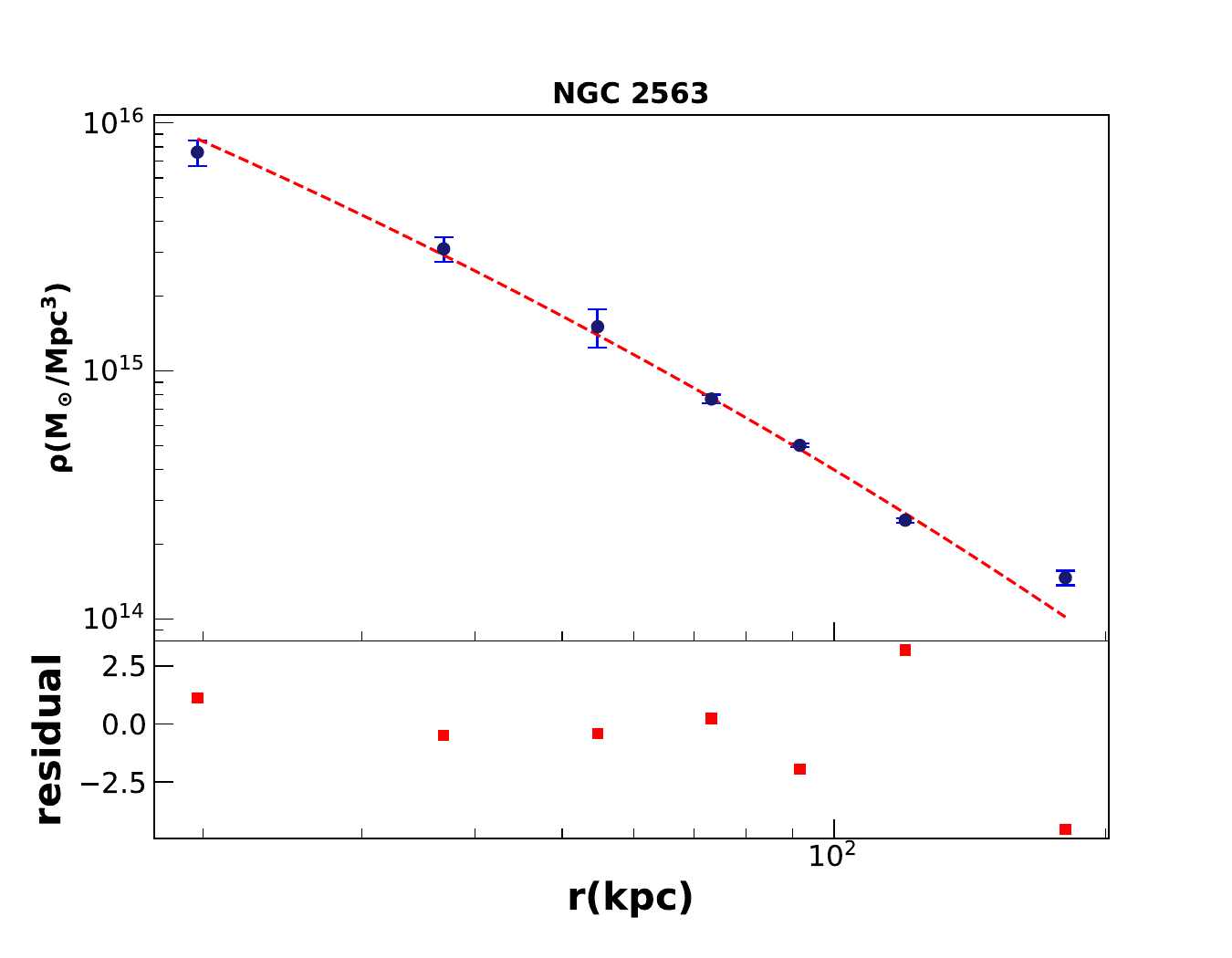}
         \label{figein:fig1.7}
     \end{subfigure}    
         \hfill
     \hfill 
                \begin{subfigure}[b]{0.45\textwidth}
         \centering
         \includegraphics[width=\textwidth]{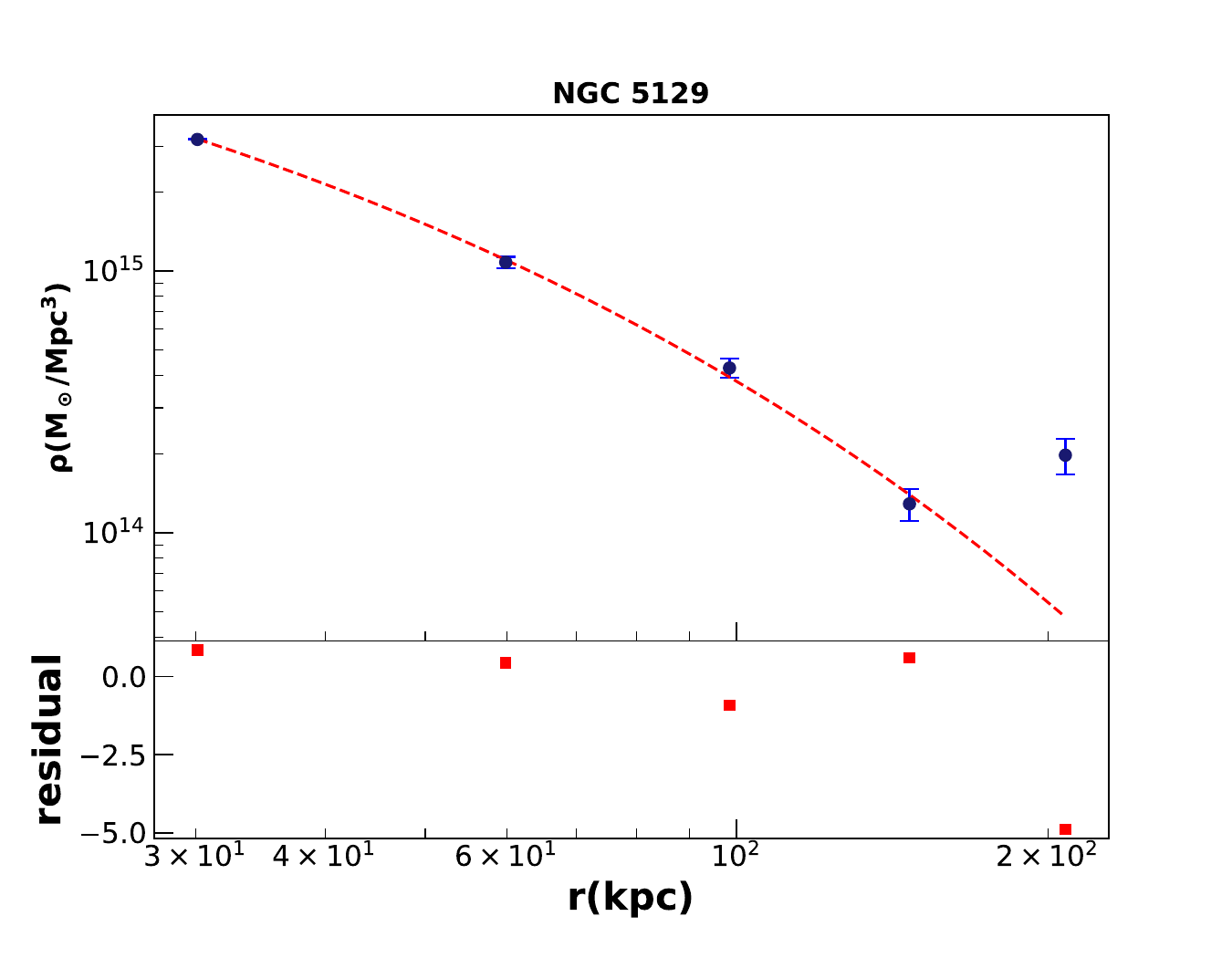}
         \label{figein:fig1.8}
     \end{subfigure}    
         \hfill
     \hfill 

     \begin{subfigure}[b]{0.45\textwidth}
         \centering
         \includegraphics[width=\textwidth]{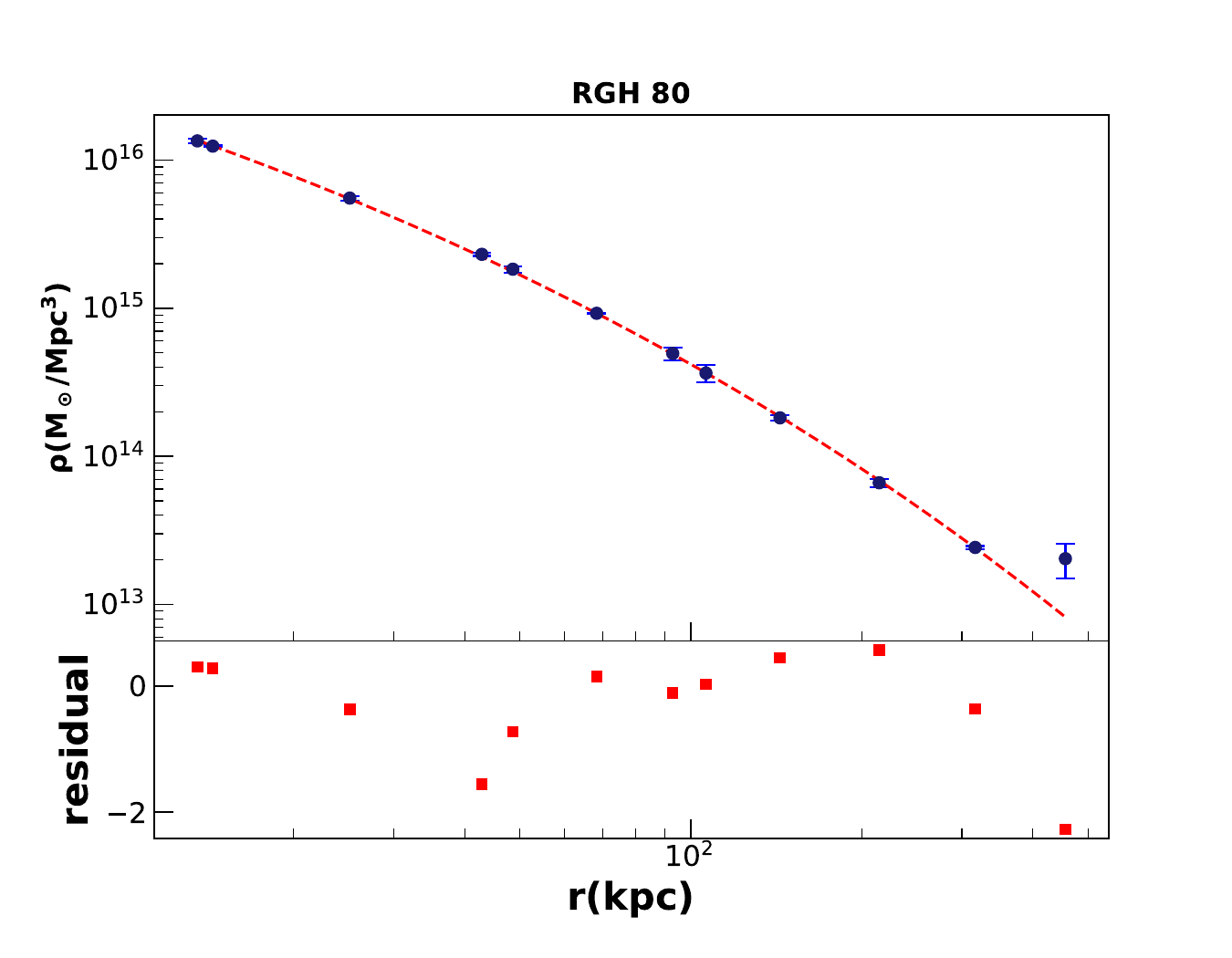}
         \label{figein:fig1.9}

     \end{subfigure}
    \hfill
     \hfill
     \begin{subfigure}[b]{0.45\textwidth}
         \centering
         \includegraphics[width=\textwidth]{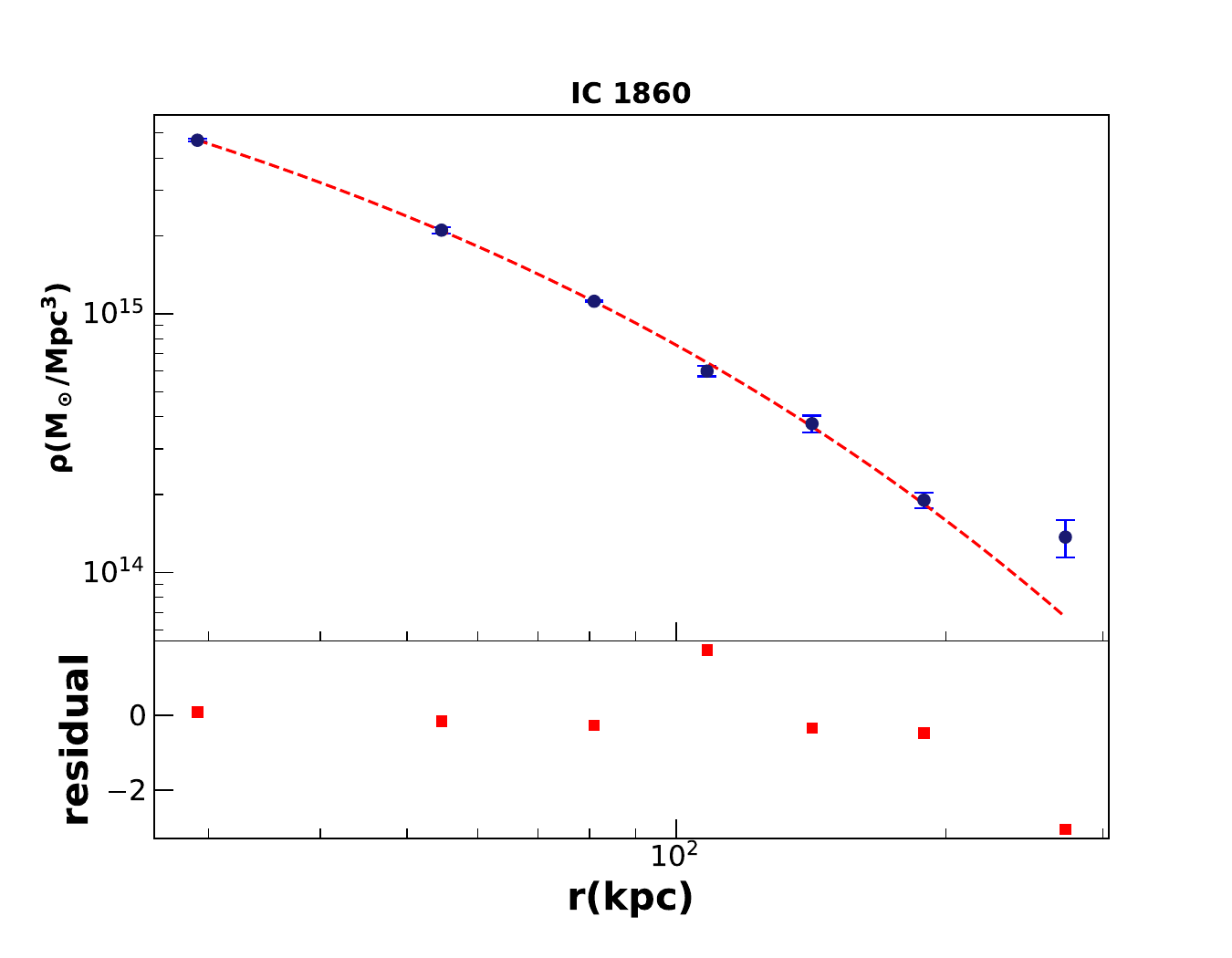}
         \label{figein:fig1.10}
     \end{subfigure}
       \hfill
     \hfill
               \begin{subfigure}[b]{0.45\textwidth}
         \centering
         \includegraphics[width=\textwidth]{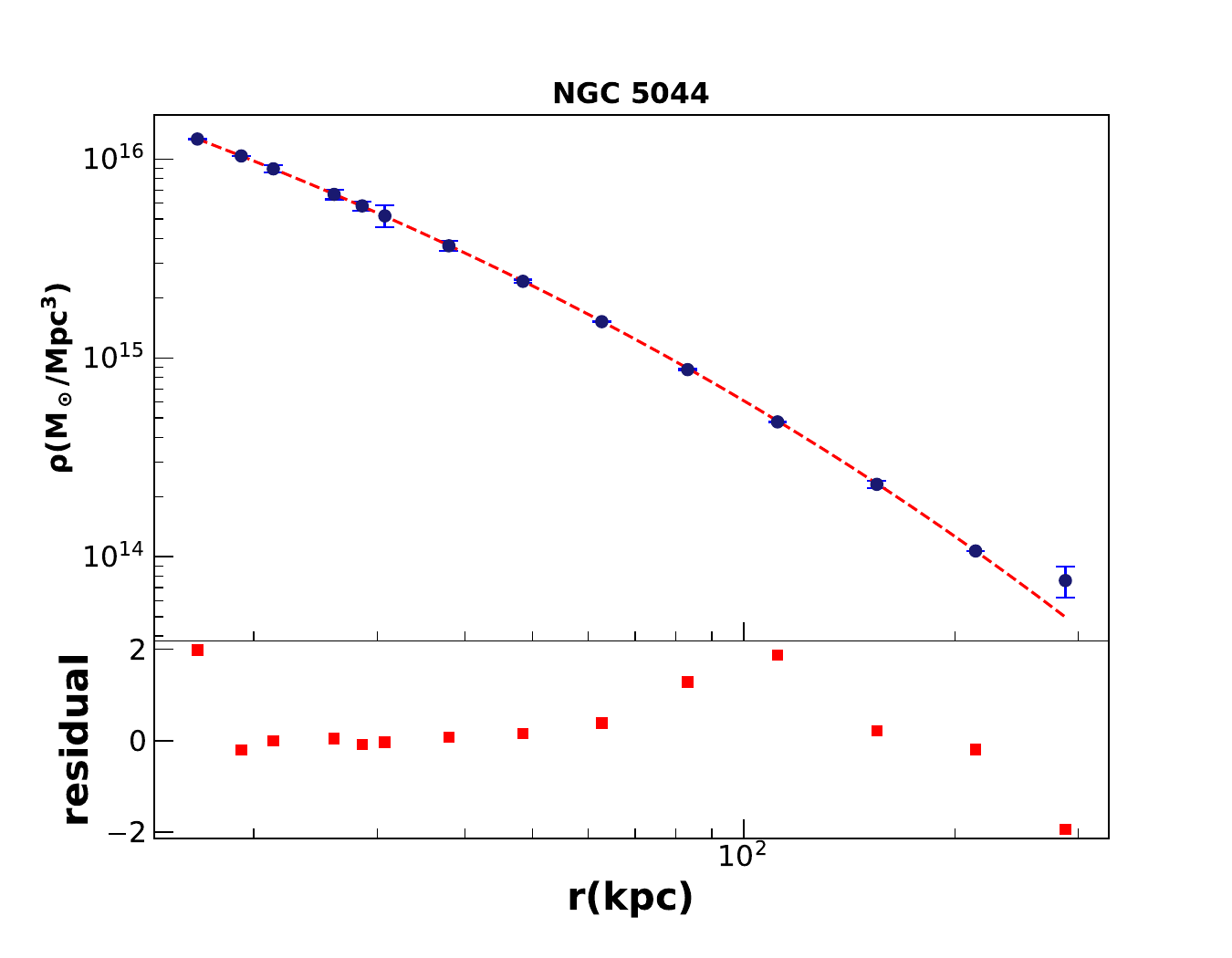}
         \label{figein:fig1.11}
     \end{subfigure}    
         \hfill
     \hfill 
            \caption{Dark matter density profile for the galaxy groups fitted with an Einasto model (contd). The plot description is the same as for the other groups in previous page.}

        \label{fig:Fig1supp}
\end{figure*}
%%%%%%%%%%%%%%%%%%%%%%%%%%%%%%%%%%%%%%%%%

\begin{figure}
     \centering
        \includegraphics[width=0.45\textwidth]{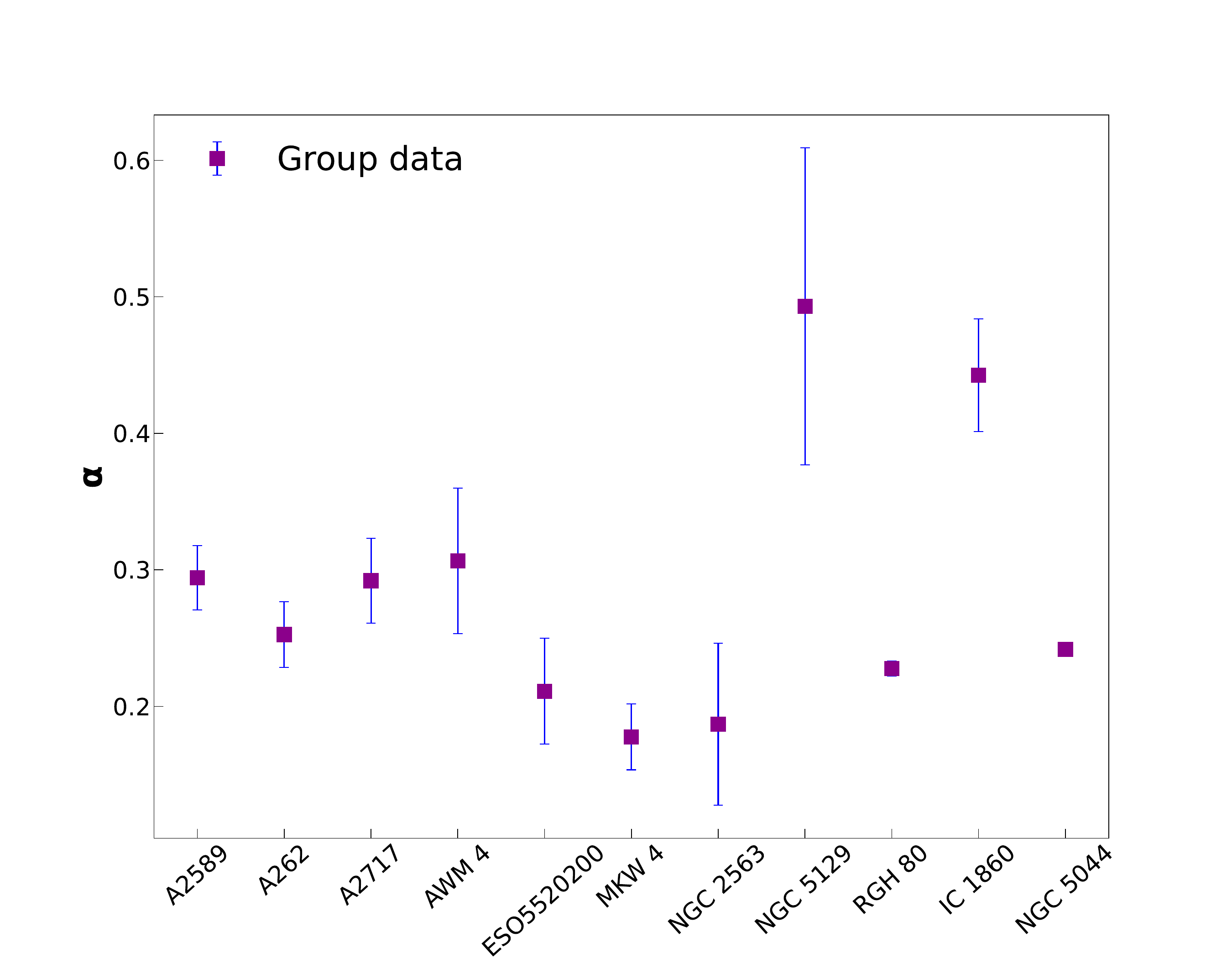}
         \label{fig:fig2}
     \hfill
     \hfill

            \caption{The Einasto shape parameter ($\alpha$) calculated for all the galaxy groups analyzed in this work.}
 \label{fig:alpha}
\end{figure}

\begin{table*}
\centering
\begin{tabular}{lcccccccccc}
\hline
\hline
\textbf{Group} && $\mathbf{\rho_s(10^{14}M_\odot Mpc^{-3})}$ && $\mathbf{r_s(kpc)}$ && $\mathbf{\alpha}$ && \textbf{DOF} && $\mathbf{\chi^2_{red}}$\\
\hline
\hline
A2589 && $6.66 \pm 0.35$ && $130.74 \pm 3.94$ && $0.29 \pm 0.02$ && 6 && 1.34 \\
A262 && $4.39 \pm 0.75$ && $161.16 \pm 17.16$ && $0.25 \pm 0.02$ && 16 && 0.002\\
A2717 && $2.22 \pm 0.25$ && $253.23 \pm 13.38$ && $0.29 \pm 0.03$ && 4 &&0.90\\ 
AWM 4 && $3.88 \pm 0.98$ && $204.42 \pm 29.99$ && $0.31 \pm 0.05$ && 5 && 0.05\\
ESO 5520200 && $2.88 \pm 0.31$ && $208.57 \pm 12.32$ && $0.21 \pm 0.04$ && 4 && 0.27 \\
MKW 4 && $17.7 \pm 2.02$ && $68.24 \pm 3.78$ && $0.18 \pm 0.02$ && 13 && 1.74\\
NGC 2563 && $11.1 \pm 3.05$ && $61.47 \pm 7.71$ && $0.19 \pm 0.06$ && 4 && 9.02\\
NGC 5129 && $7.75 \pm 1.15$ && $71.86 \pm 5.38$ && $0.49 \pm 0.12$ && 2 && 2.21\\
RGH 80 && $8.76 \pm 0.30$ && $70.24 \pm 1.22$ && $0.23 \pm 0.01$ && 9 && 1.01\\
IC 1860 && $5.60 \pm 0.75$ && $117.53 \pm 9.07$ && $0.40 \pm 0.04$ && 4 && 1.16 \\
NGC 5044 && $8.62 \pm 0.1$ && $84.42 \pm 0.18$ && $0.24 \pm 0.003$ && 11 &&  1.19 \\

\hline
\end{tabular}
\caption{\label{tab:Einasto} This table shows the best-fit parameter values for the Einasto profile (cf. Eq.~\ref{eq:einasto}) to the group halos along with the degrees of freedom (DOF). The efficacy of the fit can be quantified by the reduced $\chi^2$ (shown in the last column).  The data for A2589 was obtained from ~\cite{Zappacosta_2006}  and for all  other groups from ~\cite{Gastaldello_2007}.}
\end{table*}

\subsection{Results for  $\sigma/m$}
Once we have determined the best-fit $\alpha$ for each group, we  obtain an estimate for $\sigma/m$ by comparing with Eq.~\ref{eq:alphasidm}. We assume that this equation is also valid for low mass clusters and groups with masses in the range $10^{13}-10^{14} M_{\odot}$, as no new physics kicks in below $10^{14} M_{\odot}$. Furthermore, our current sample also contains five objects with $M_{vir}>10^{14} M_{\odot}$. If the observed  $\alpha$ for a given group is consistent  (within errors) we  set an upper limit on  $\sigma/m$. Else one could get a   central bounded estimate for $\sigma/m$. For this purpose, we calculate $\chi^2$ as a function of $\sigma/m$ as follows:
\begin{equation}
    \chi^2 = \left[\frac{\alpha_{obs} - \alpha_{th}(\alpha_0,\alpha_1,\gamma,\sigma)}{\sigma_{\alpha}}\right]^2,
    \label{eq:chialpha}
\end{equation}
where $\alpha_{th}(\alpha_0,\alpha_1,\gamma,\sigma)$ is given by Eq.~\ref{eq:alphasidm}, $\alpha_{obs}$ is the observed value for each group and $\sigma_{\alpha}$ its associated error.
If the $\chi^2$ functional is shaped like a parabola   with a well-defined minimum, one could get bounded c.l. intervals for $\sigma/m$ based on fixed $\Delta \chi^2$ values compared to the minimum~\cite{NR}, as long as the lower point of intersection is greater than 0.  

%In case the reduced $\chi^2>1$, there are many examples throughout astrophysics, cosmology and particle physics, where an additional systematic error has been added either in quadrature  to the observed statistical errors   or a multiplicative scale factor to the observed statistical errors, so that the reduced $\chi^2$ is equal to 1~\cite{PDG,Desai12,Pratik,Erler}. 
\rthis{For the cases when the reduced $\chi^2 > 1$ in the estimation of the SIDM cross-section, we assume that there are unaccounted for systematic effects, and rescale $\sigma_{\alpha}$ (in Eq.~\ref{eq:chialpha}) by $\chi^2_{red}$ following Ref.~\cite{PDG,Desai12,Pratik,Erler}.}\footnote{This procedure has also been criticized in literature~\cite{Sarkar}. In the case of  linear regression, one can show that the errors in the estimated parameters  need to be rescaled by  $\sqrt{\chi^2_{red}}$~\cite{NR}. Although such a rescaling  may  not be exact  for non-linear regression or if the  errors are not Gaussian, this  rescaling procedure has been applied in a number of studies  from Particle Physics (where it is routinely used by Particle Data Group to rescale the errors in the masses or lifetimes of elementary particles~\cite{PDG,Erler}) to pulsar timing, (where the errors in the  pulsar times of arrival  have been rescaled based on reduced $\chi^2$~\cite{Pratik}.)}
Therefore, for our analysis,  in case the reduced $\chi^2$ for any group is greater than 1, for estimating the SIDM cross-section, we rescaled the  $\sigma_{\alpha}$ (in Eq.~\ref{eq:chialpha}) by $\sqrt{\chi^2_{red}}$. The  $\chi^2$ curves  as a function of $\sigma/m$ are  shown in Fig.~\ref{fig:chivsalpha}. We have a total of two groups with reduced $\chi^2 > 2$. So this rescaling would at most drastically affect the results of only two groups.

Since all other works in literature have obtained limits on SIDM at 95\% c.l, we also obtain limits (or central estimates) at 95\% c.l. These are obtained by  finding the maximum  value of $\sigma/m$ for which $\Delta \chi^2 < 4$~\cite{NR}. In case the $\chi^2$ curve  shows a minimum for $\sigma/m >0$,  we  report 95\% central estimate  if the lower value of $\sigma/m$ for which $\Delta \chi^2=4$ is greater than 0.     A comprehensive summary of our results  for  the 11 groups  can be found in Table~\ref{table:results}. We find non-zero central estimates for $\sigma$/m for seven groups, whereas for the remaining four groups we report upper limits at 95\% c.l. However among these  seven groups, only one group, viz. NGC 5044 has fractional error  $<$ 20\%, given by $\sigma/m=0.165 \pm 0.025~\rm{cm^2/g}$. The upper limits on $\sigma/m$ which we obtain for the remaining  groups are between $0.16-6.61~\rm{cm^2/g}$. The most stringent limit is obtained for the group MKW 4, which has $\sigma/m<0.16~\rm{cm^2/g}$. 

In addition to estimating the limits (or central intervals) on the SIDM cross-section, we have also estimated the dark matter  collision velocity for each of the groups, so that a  comparison can be made with the currently viable models of SIDM cross-section decreasing with velocity~\cite{SIDM}. For this purpose,  we used the following scaling relation between line of sight velocity dispersion of galaxies ($\sigma_{LOS}$) in the cluster and $M_{200}$ from ~\cite{Munari}, (which was also used in ~\cite{Sagunski}):
\begin{equation}
\log (h(z) M_{200})=13.98 + 2.75 \log \left[\frac{\sigma_{LOS}}{500~\rm{(km/sec)}}\right]
\label{eq:munari}
\end{equation}
To estimate $\sigma_{LOS}$ and its uncertainty, we used the values of $M_{200}$ and its error from ~\cite{Gastaldello_2007,Zappacosta_2006}. We also note that there are also variations of order 10\% reported in ~\cite{Munari} in the exponent and slope in Eq.~\ref{eq:munari}, depending on the simulation set used. We do not take this into account while estimating the uncertainty. However, this uncertainty would uniformly affect all the groups. These values of $\sigma_{LOS}$ can be found in Table~\ref{table:results}.  Therefore, the groups with  95\% upper limits between $0.16-6.61~\rm{cm^2/g}$ have dark matter collision velocity between approximately 200-500 km/sec with the most stringent limit for MKW 4 ($\sigma/m <0.16 ~\rm{cm^2/g}$) at a velocity dispersion of $\approx$ 350 km/sec.
 Conversely,  the group with the most precise non-zero estimate for $\sigma$/m, viz. NGC 5044 has dark matter velocity dispersion of about 300 km/sec.

 Therefore, our results are  consistent with the constraints on SIDM using galaxy groups obtained in ~\cite{Sagunski} and agree  with predictions of  velocity-dependent SIDM cross-sections on group/cluster scales~\cite{Kaplinghat15}. The constraints on $\sigma/m$, which we obtain are  within the ballpark of $0.1-1~\rm{cm^2/g}$ needed to solve the core-cusp and too big to fail problems~\cite{Tulin}. 

\begin{figure*}
     \centering

     \begin{subfigure}[b]{0.4\textwidth}
         \centering
         \includegraphics[width=\textwidth]{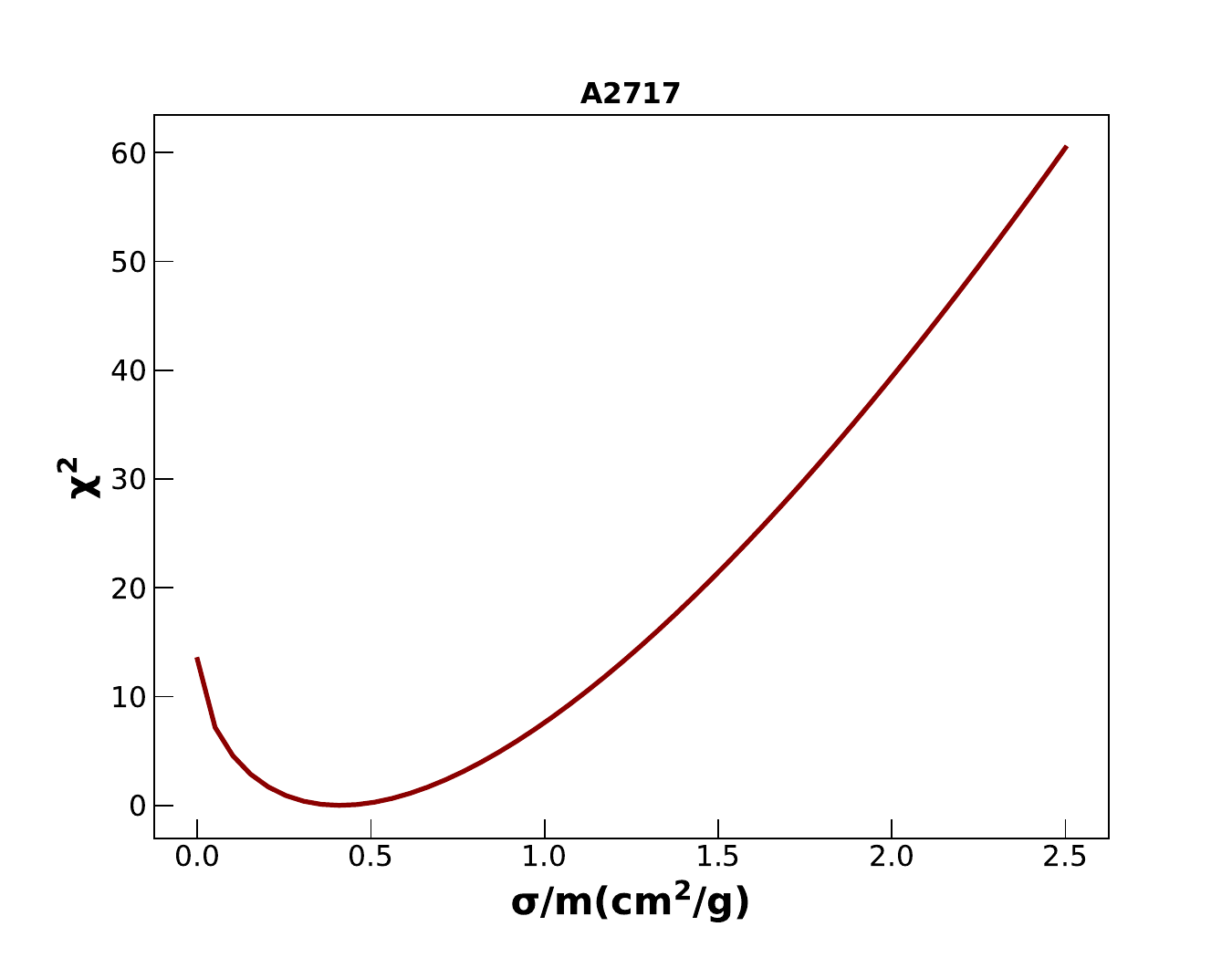}
         \label{fig:f2.2}
     \end{subfigure}
     \hfill
     \hfill
     \begin{subfigure}[b]{0.4\textwidth}
         \centering
         \includegraphics[width=\textwidth]{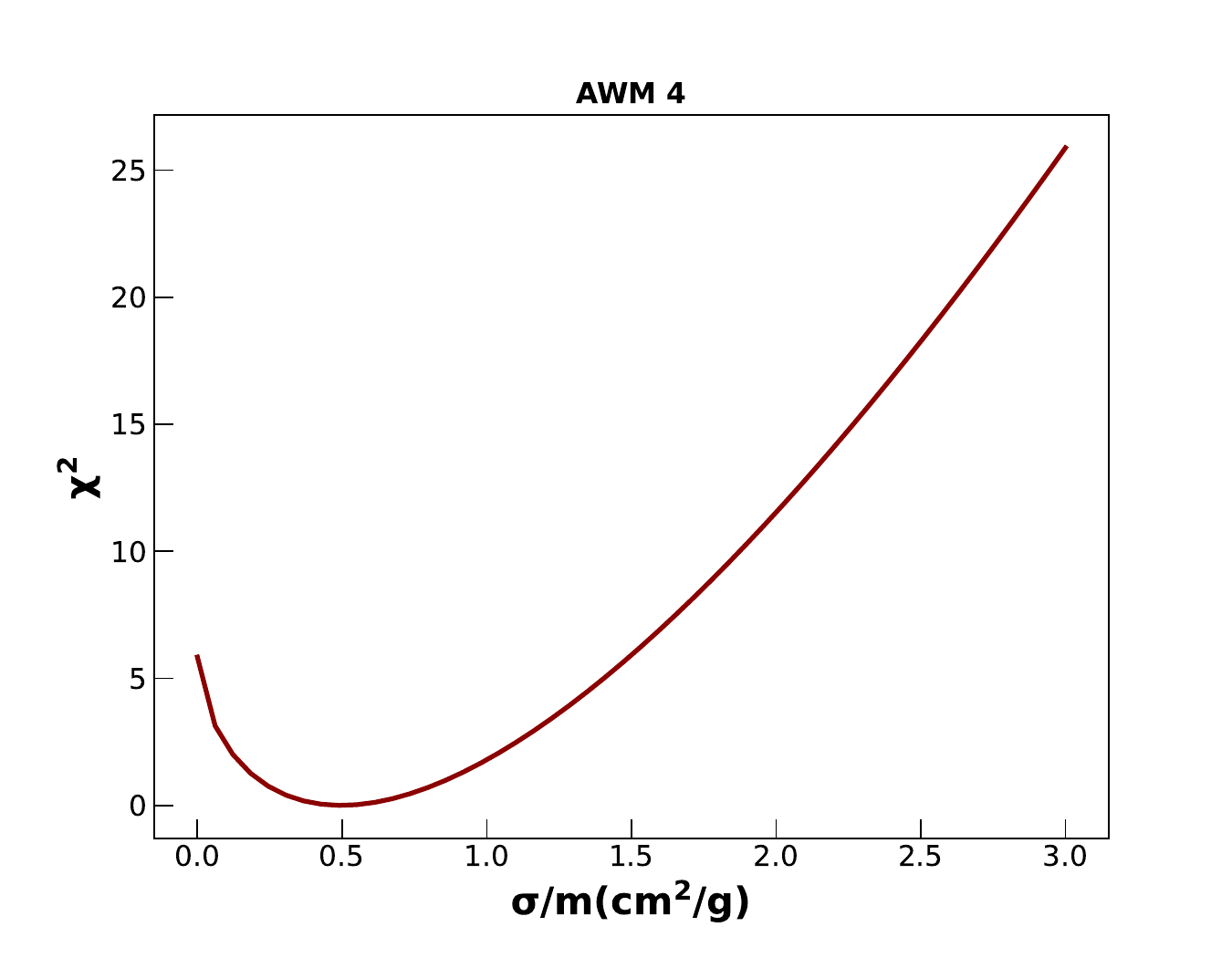}
         \label{fig:f2.3}
     \end{subfigure}
     \hfill
     \hfill
      \begin{subfigure}[b]{0.4\textwidth}
         \centering
         \includegraphics[width=\textwidth]{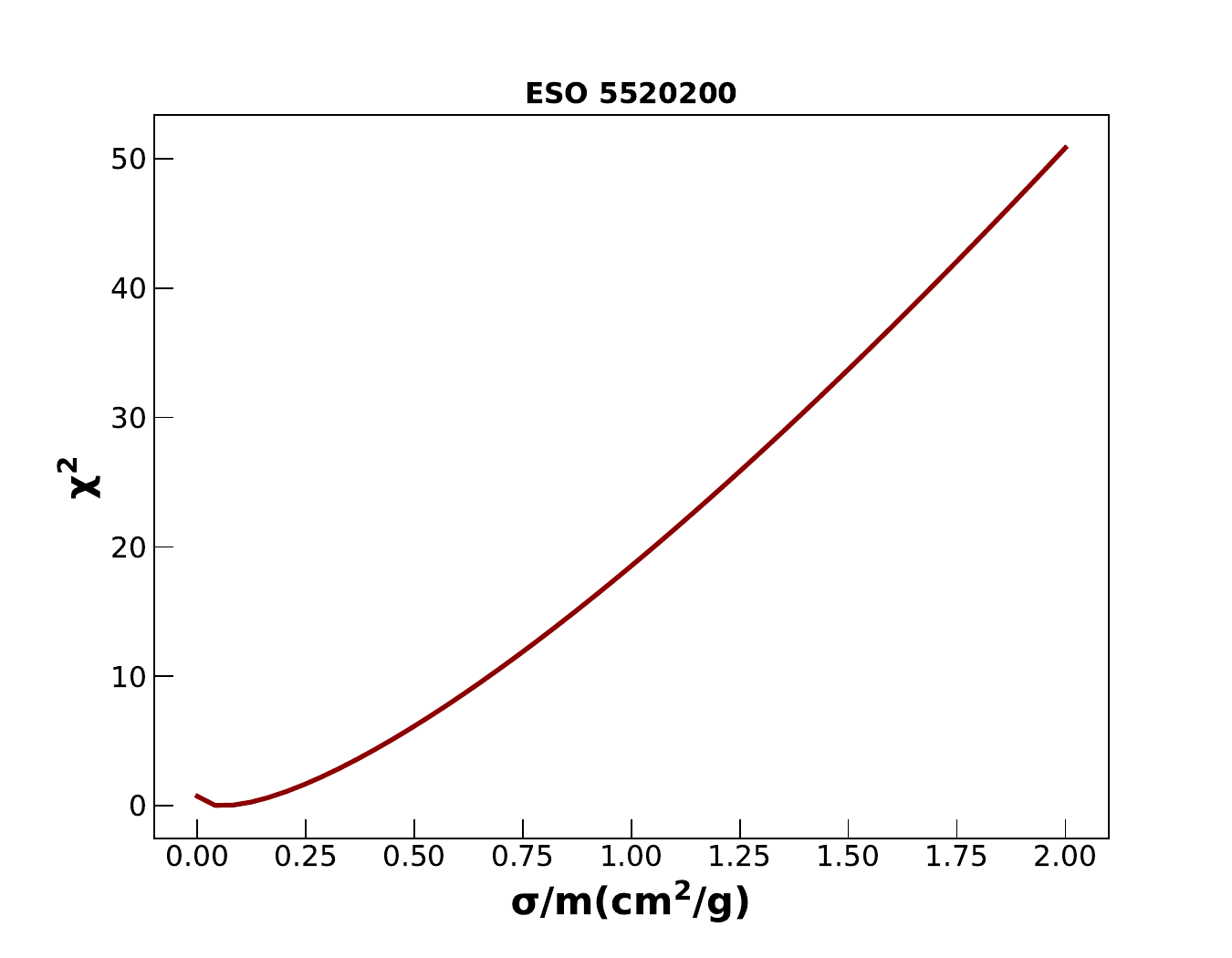}
         \label{fig:f2.4}
     \end{subfigure}    
         \hfill
     \hfill   
          \begin{subfigure}[b]{0.4\textwidth}
         \centering
         \includegraphics[width=\textwidth]{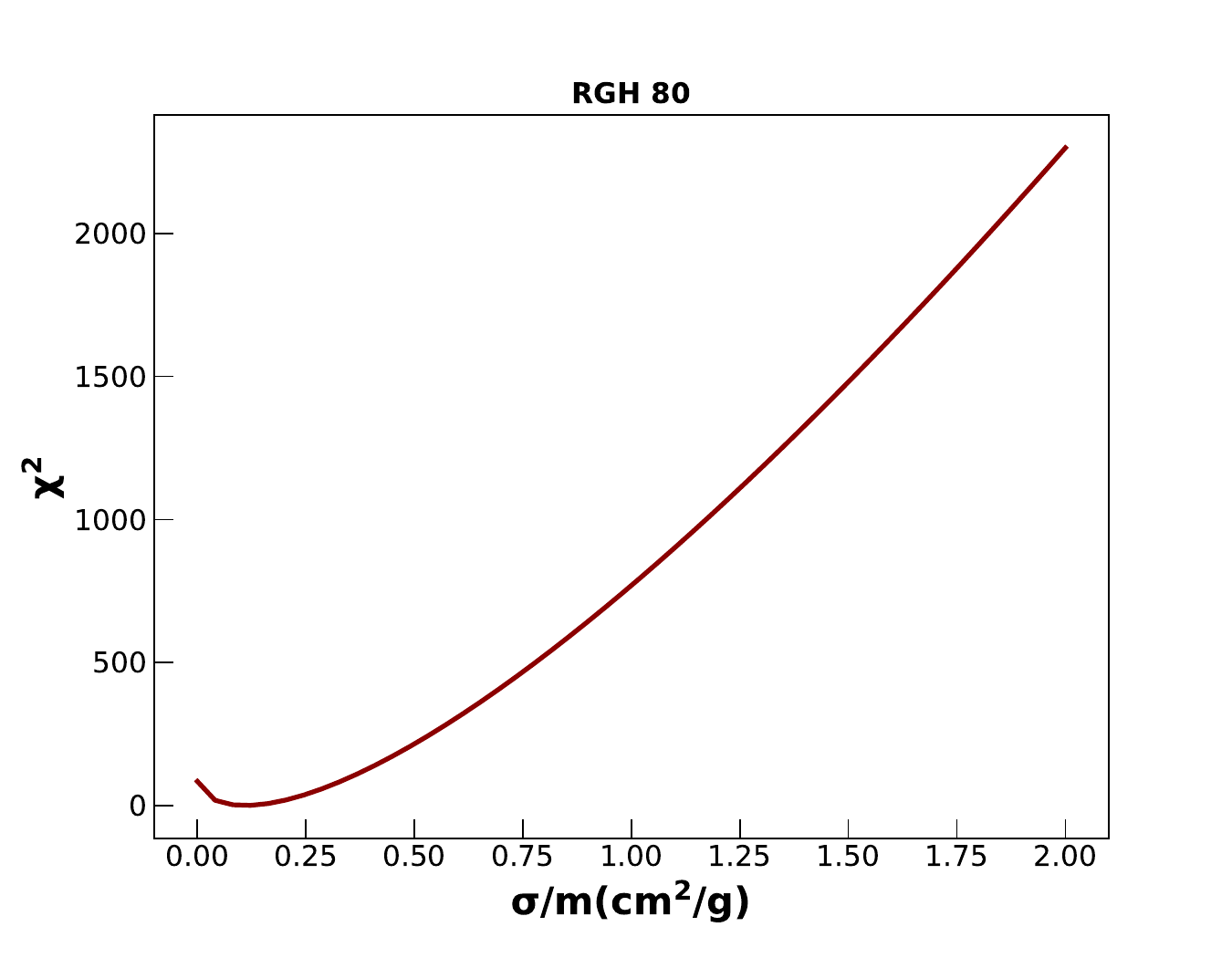}
         \label{fig:f2.5}
     \end{subfigure}
       \hfill
     \hfill
     \begin{subfigure}[b]{0.4\textwidth}
         \centering
         \includegraphics[width=\textwidth]{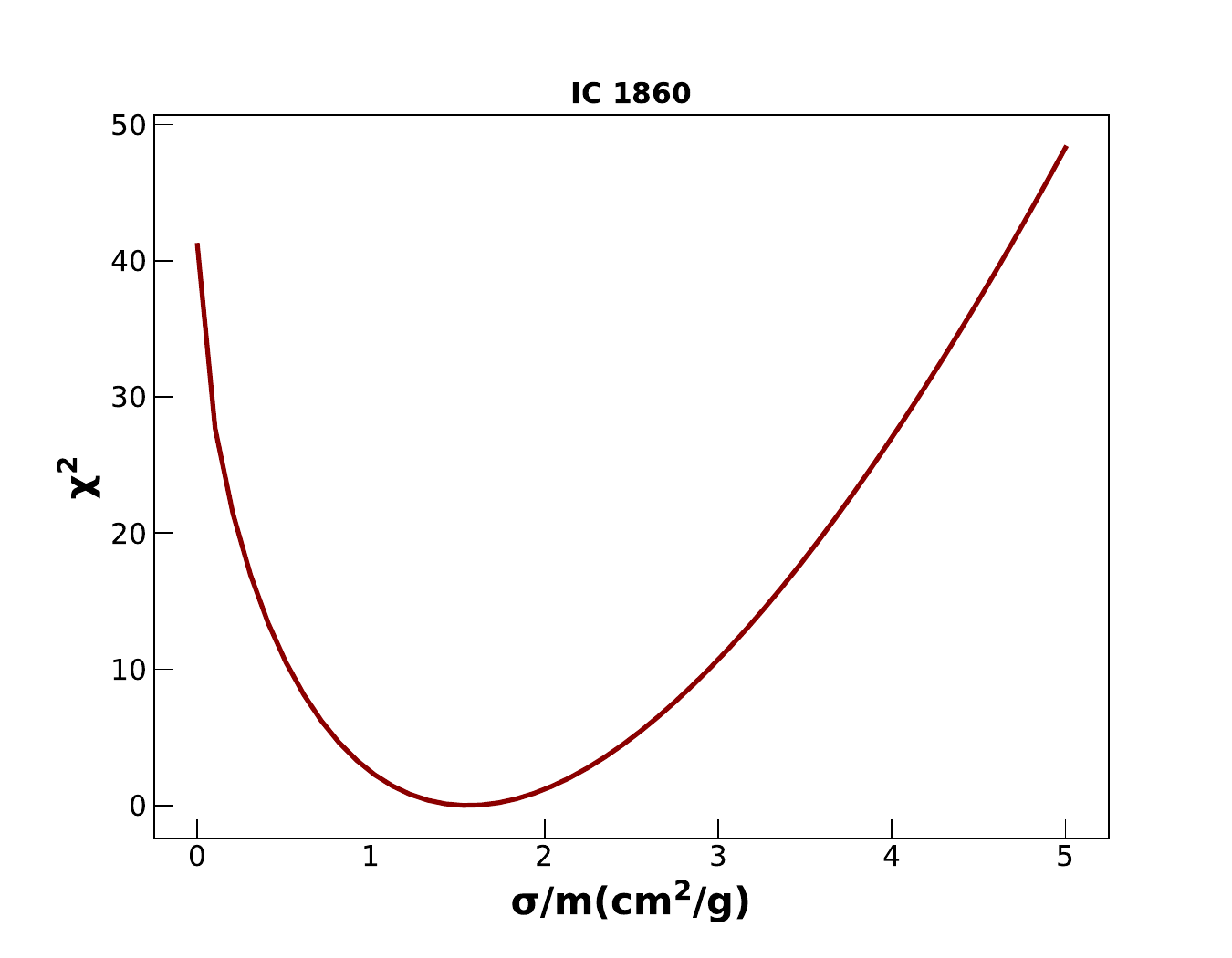}
         \label{fig:f2.6}
     \end{subfigure}
     \hfill
     \hfill
     \begin{subfigure}[b]{0.4\textwidth}
         \centering
         \includegraphics[width=\textwidth]{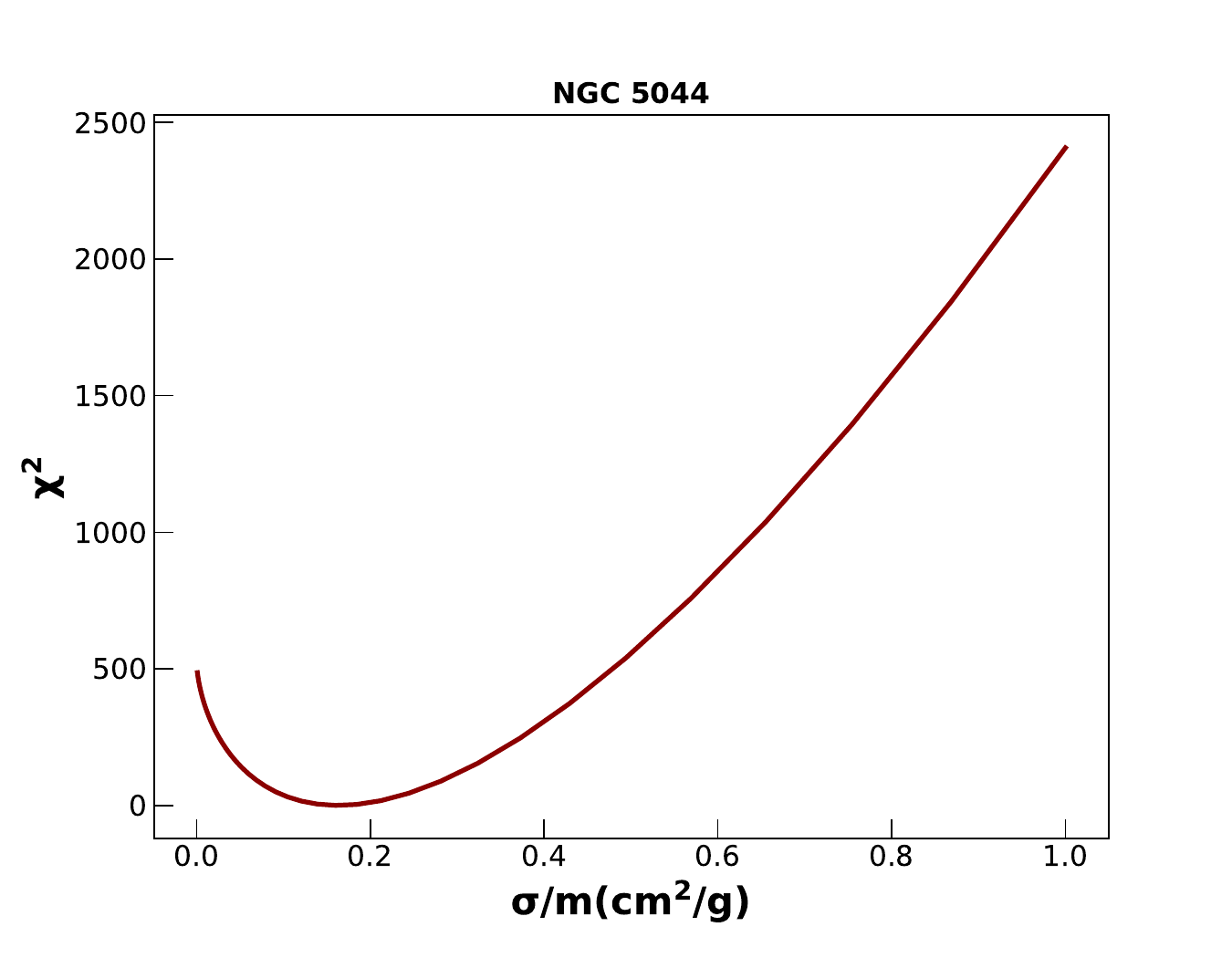}
         \label{fig:f2.7}
     \end{subfigure}
     \hfill
     \hfill
               \begin{subfigure}[b]{0.4\textwidth}
         \centering
         \includegraphics[width=\textwidth]{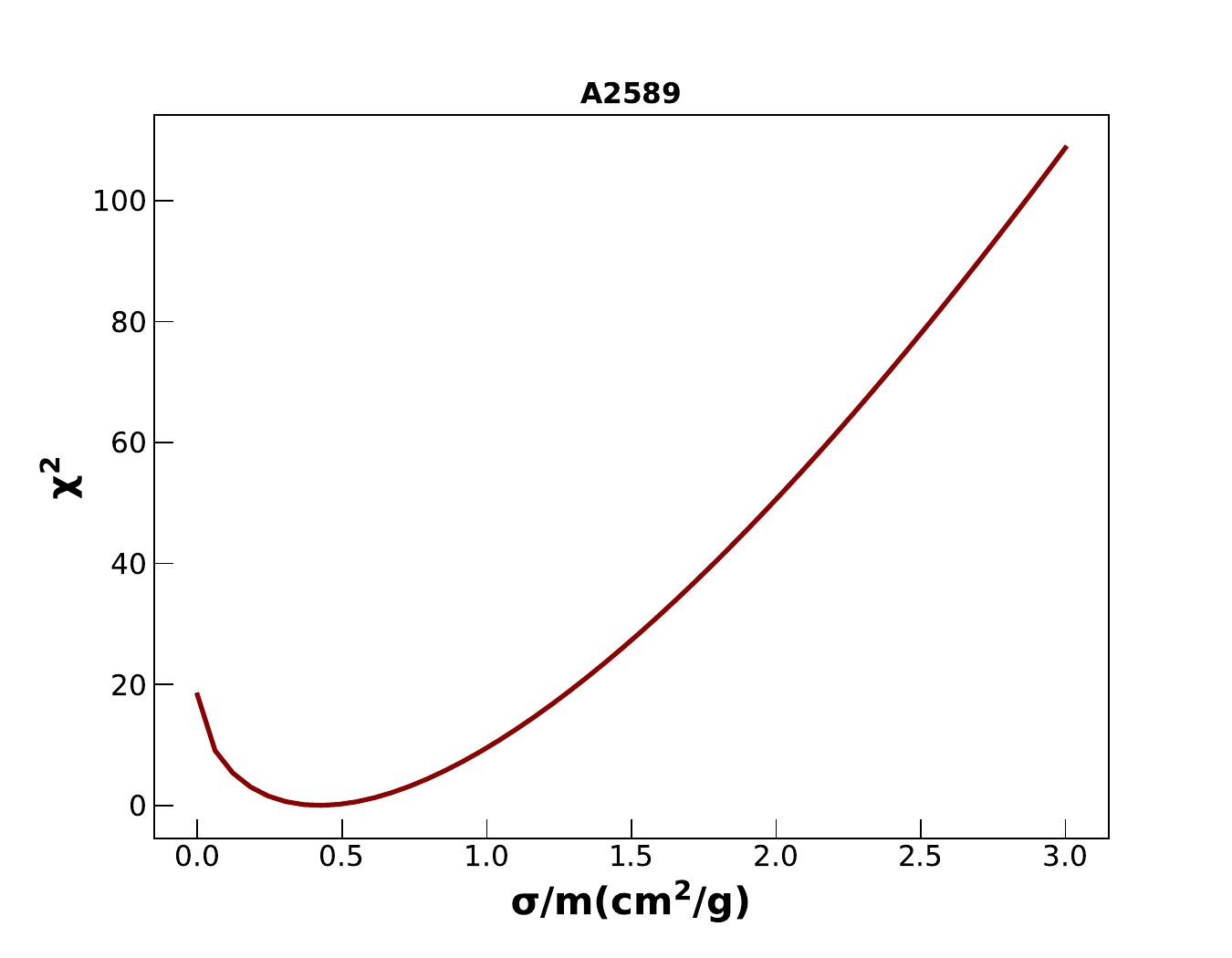}
         \label{fig:f2.8}
     \end{subfigure}
     \hfill
     \hfill
     \begin{subfigure}[b]{0.4\textwidth}
         \centering
         \includegraphics[width=\textwidth]{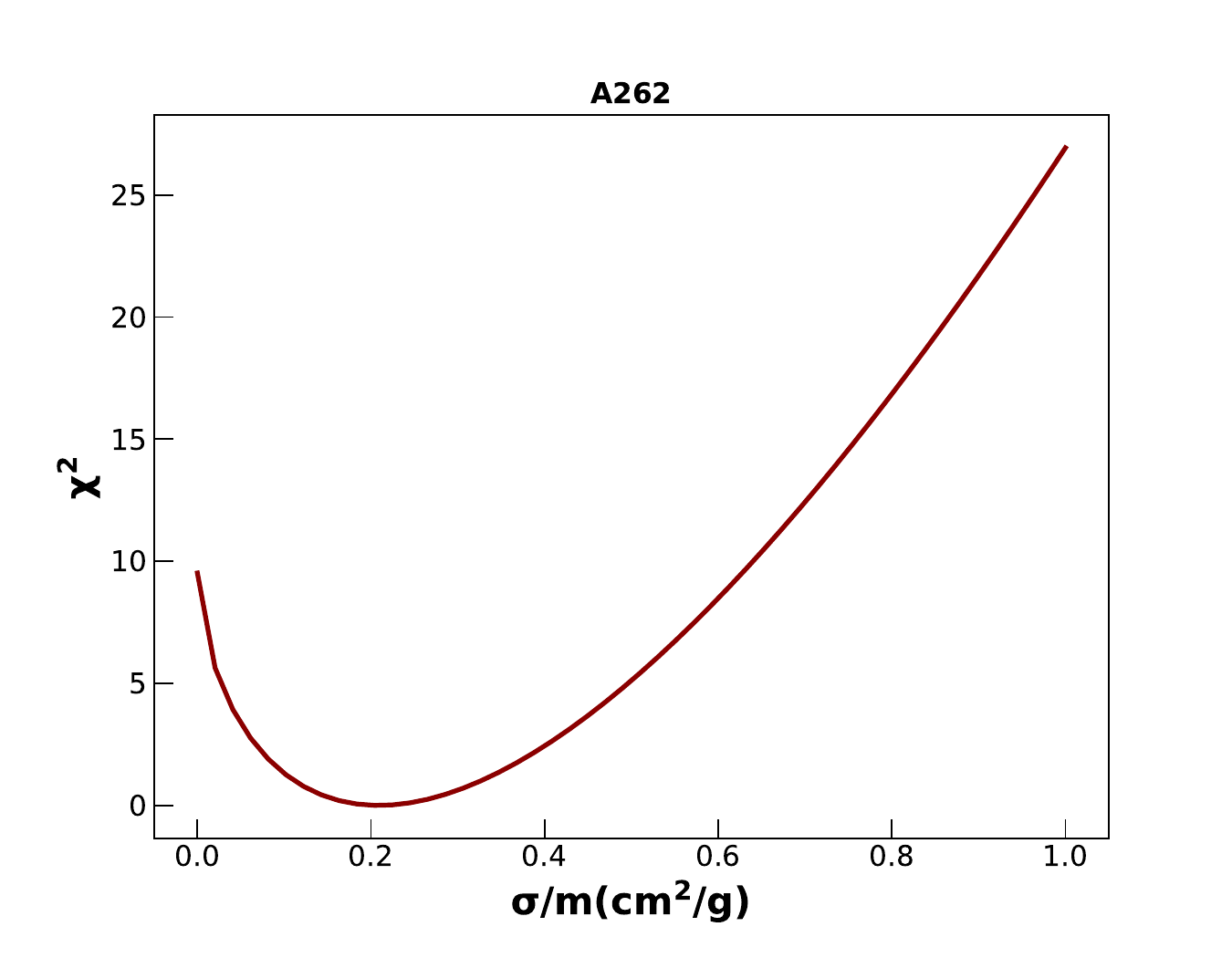}
         \label{fig:f2.1}
     \end{subfigure}
        \hfill
          \hfill

            \caption{Distribution of $\chi^2$ as a function of the dark matter self-interaction cross-section $\sigma/m$ for all the 11 groups in our sample. }
 \label{fig:chivsalpha}
\end{figure*}

\begin{figure*}\ContinuedFloat
     \centering

     \begin{subfigure}[b]{0.4\textwidth}
         \centering
         \includegraphics[width=\textwidth]{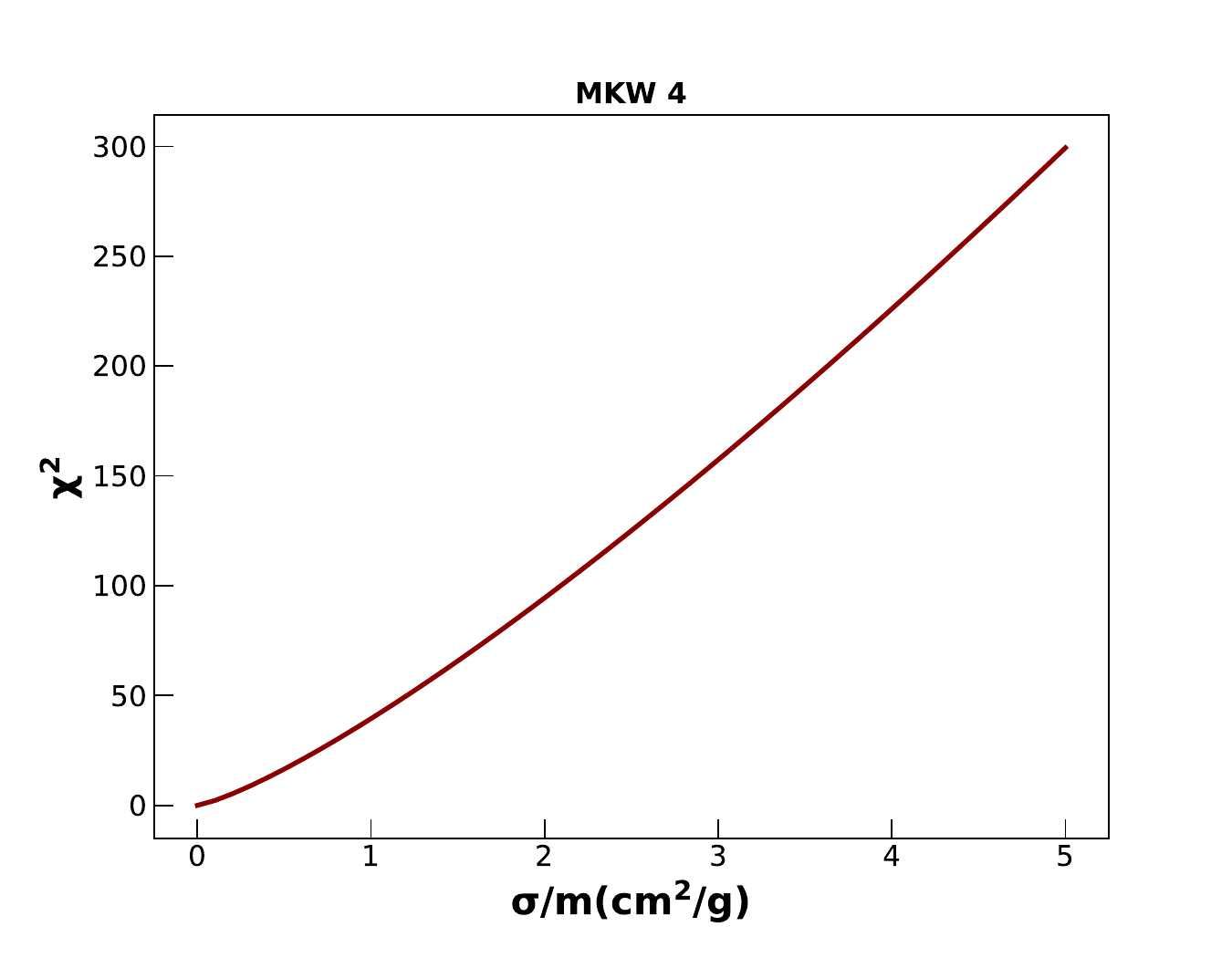}
         \label{fig:f2.9}
     \end{subfigure}
     \hfill
     \hfill
          \begin{subfigure}[b]{0.4\textwidth}
         \centering
         \includegraphics[width=\textwidth]{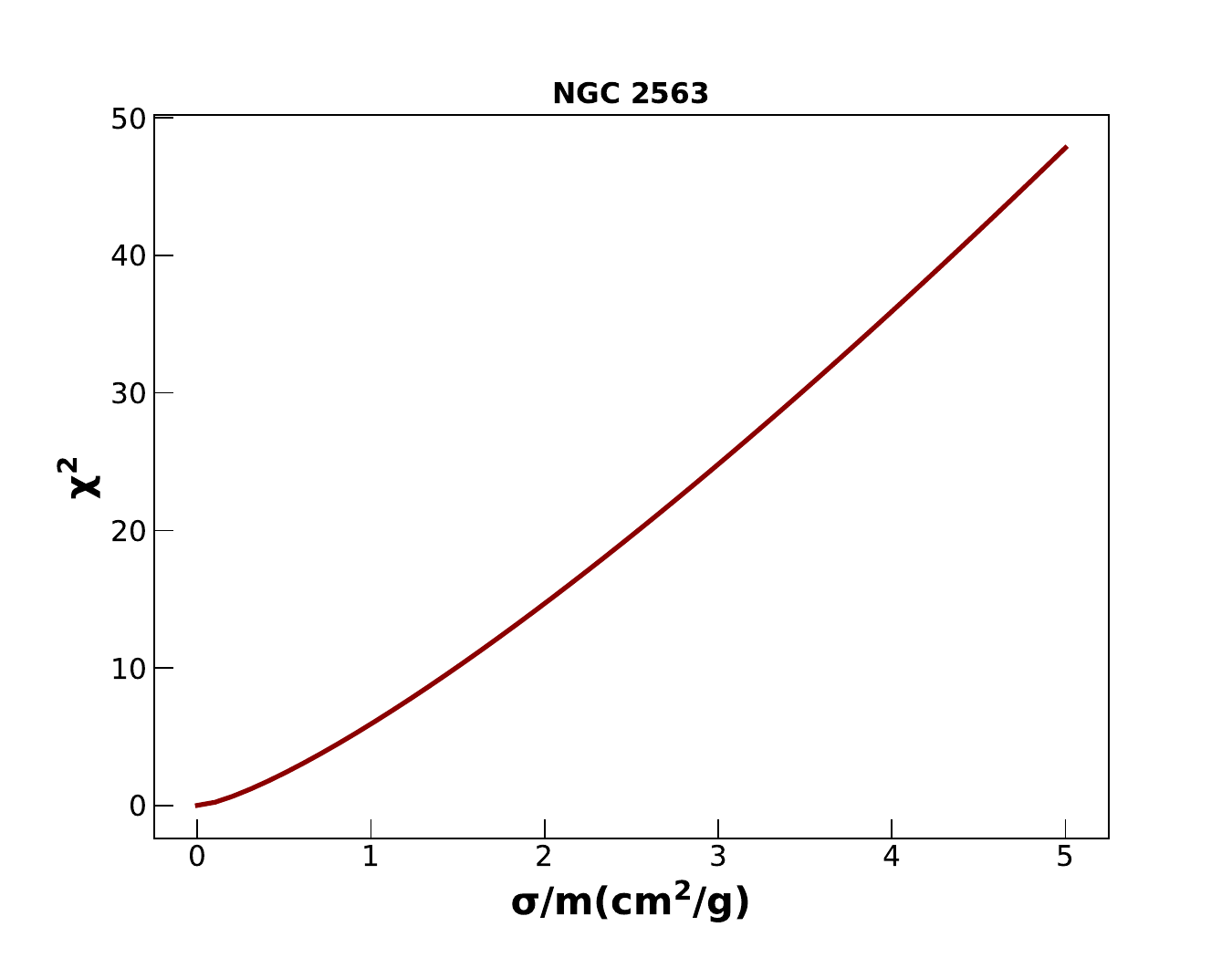}
         \label{fig:f2.10}
     \end{subfigure}
     \hfill
     \hfill
          \begin{subfigure}[b]{0.4\textwidth}
         \centering
         \includegraphics[width=\textwidth]{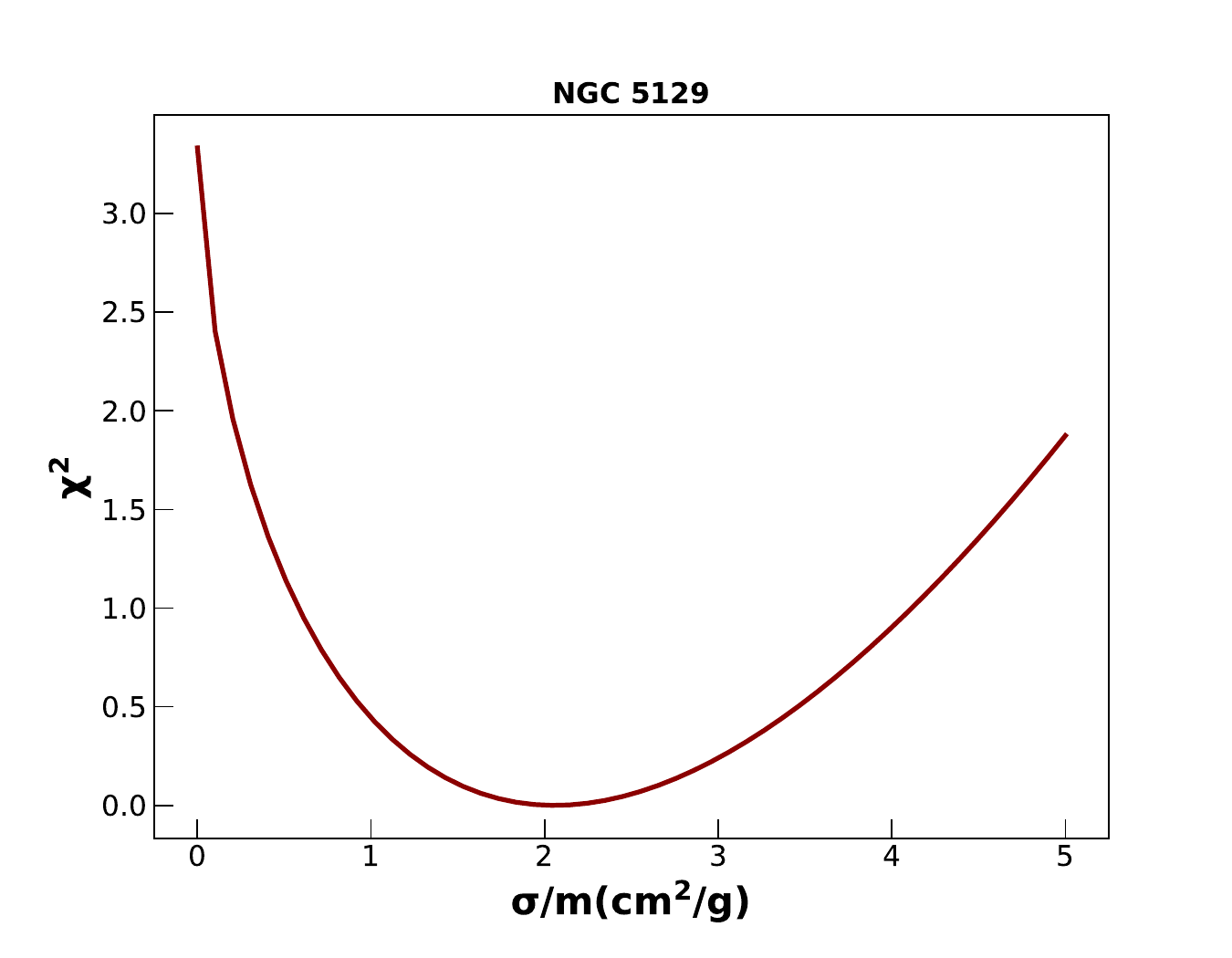}
         \label{fig:f2.11}
     \end{subfigure}
     \hfill
     \hfill
            \caption{Distribution of $\chi^2$ as a function of the dark matter self-interaction cross-section $\sigma/m$ for groups (contd.) }
 \label{fig:chivsalpha2}
\end{figure*}

\begin{table}[h]
\centering
\begin{tabular}{|c|c|c|}
\hline
\textbf{Group} & \textbf{Limit (Estimate) on $\sigma/m$} & 
\textbf{$\sigma_{LOS}$}\\
& $\rm{(cm^2/g})$ & (km/sec) \\
\hline
A262 & $0.25 \pm 0.21$ & 436 $\pm$ 32  \\
A2717 & $0.47 \pm 0.35$  &  523 $\pm$ 26 \\
AWM 4 & $0.67 \pm 0.63$ & 504 $\pm$ 45\\
ESO 5520200 & $<0.39$ & 463 $\pm$ 54 \\
RGH 80 & $0.11 \pm 0.038 $  & 267 $\pm$ 11 \\
IC 1860 & $1.63 \pm 0.87 $  &  329 $\pm$ 23\\
NGC 5044 & $0.165 \pm 0.025$  &  295 $\pm$ 8 \\
A2589 & $0.46 \pm 0.31$  &  694 $\pm$ 53 \\
MKW 4 & $<0.16$  &  358 $\pm$ 15 \\
NGC 2563 & $<2.61$  &  259 $\pm$ 42 \\
NGC 5129 & $<6.61$  &  216 $\pm$ 24 \\
\hline
\end{tabular}
\caption{\label{table:results} 95\% c.l. upper limits (or central intervals) on SIDM cross-section for the galaxy groups considered in this work along with the line of sight velocity dispersion of galaxies obtained using the scaling relation in Eq.~\ref{eq:munari}~\cite{Munari} and based on the $M_{200}$ values in ~\cite{Gastaldello_2007,Zappacosta_2006}.  For all galaxy groups with reduced $\chi^2>1$ the upper limits were determined by re-scaling the errors in $\alpha$ by $\sqrt{\chi^2_{red}}$.}
\end{table}

\section{Conclusions}
\label{sec:conclusions}
In this work, we obtain constraints on the  SIDM cross-sections using a sample of  11 relaxed galaxy groups with X-ray observations, which  we have previously used to test the invariance of the dark matter halo surface density and RAR~\cite{Gopika2021}. For this purpose, we follow the same prescription as E22, that derived  an empirical relation between the  SIDM cross-selection ($\sigma/m$) and the $\alpha$ parameter of the Einasto profile using simulated SIDM cluster halos from the {\tt Bahamas-SIDM} set of simulations~\cite{Eckert22}. This empirical relation between the Einasto $\alpha$ parameter and the SIDM $\sigma/m$ can be found in Eq.~\ref{eq:alphasidm}.

We tried to fit the density profiles of 17 galaxy groups to the Einasto parameterization.
We were able to obtain closed contours for the Einasto parameters for 11 of these groups.
The best-fit Einasto parameters  for all these groups along with the reduced $\chi^2$  can be found in Table~\ref{tab:Einasto}. 
The values of $\alpha$ which we get are between 0.12-0.49.   To obtain constraints on the SIDM cross section,  we used the aforementioned 11 groups and rescaled the errors in $\alpha$ by $\sqrt{\chi^2_{red}}$ for the groups with $\chi^2_{red}>1$. Possible reasons for the large reduced $\chi^2$ for some of these groups could be due to AGN feedback in the core of the group or incomplete relaxation~\cite{Gastaldello_2007}.
We then obtain the 95\% c.l. upper limits (or central estimates) on the SIDM cross-section, by finding the maximum value of $\sigma/m$ for which $\Delta \chi^2=4$, where $\chi^2$ has been obtained using Eq.~\ref{eq:chialpha}.

Our results for SIDM cross-sections for the 11 relaxed groups can be found in Table~\ref{table:results}.  We obtain a non-zero value for seven groups, with the most precise estimate (fractional error $<$ 20\%)
for $\sigma/m$ for one group (NGC 5044), given by $\sigma/m=0.165 \pm 0.025~\rm{cm^2/g}$ at 95\% c.l. at a velocity dispersion of about 300 km/sec.  For the remaining four other groups, we estimate 95\% c.l. upper limits on $\sigma/m$ in the range of $0.16-6.61~\rm{cm^2/g}$ with dark matter velocity dispersions in the range of 200-500 km/sec. The most stringent limit which we obtain is for MKW 4  for which $\sigma/m < 0.16 ~\rm{cm^2/g}$ at velocity dispersion of $\approx$ 350 km/sec.   Our results  for $\sigma/m $ for all  the groups in  our sample are in the same ballpark as those found  in ~\cite{Sagunski}, (which were consistent with  $\sigma/m = 0.5 \pm 0.2~\rm{cm^2/g}$ or 
 $\sigma/m < 0.9~\rm{cm^2/g}$, if interpreted as an upper limit). Our results are also consistent  with the  predictions of velocity-dependent SIDM cross-sections at group/cluster scales~\cite{Kaplinghat15}.

\section*{Acknowledgements}
We are grateful to Fabio Gastaldello and Dominique Eckert  for useful correspondence  related to the galaxy group data used for this work and E22, respectively. We also thank the anonymous referee for several useful and constructive comments on the manuscript. GK also acknowledges the Ministry of Education (MoE), Government of India for the Senior Research Fellowship.

\bibliography{new}

\appendix

\section{Einasto fits to Galaxy Groups}
\label{sec:appendix}
In this section, we shall discuss the fitting of the Einasto profiles to the dark matter profiles derived from the X-ray brightness and temperature data. The Einasto model was fitted to the group density profiles by maximizing the log-likelihood function given below:

\begin{equation}
\begin{aligned}
\ln \mathcal{L} = -0.5\sum \bigg[\bigg(\frac{\rho_{Einasto}-\rho_{data}}{\sigma} \bigg)^2 + \ln(\sigma^2)\bigg], 
\end{aligned}
\label{eq:maxlike}
\end{equation}
where $\rho_{data}$ and $\sigma$ denote the group density profiles and their errors, respectively.

The MCMC corner plots displaying the posterior distributions with 68\%, 90\% and 99\% credible intervals of the free parameters for the Einasto model, viz $r_s$,  $\rho_s$, and  $\alpha$ are shown in Figure \ref{fig:appendix1}. The posterior distribution medians and their 1-$\sigma$ uncertainties are represented by vertical dashed lines in the histograms shown diagonally in the corner plots.

\begin{figure*}
     \centering
     \begin{subfigure}[b]{0.48\textwidth}
         \centering
         \includegraphics[width=\textwidth]{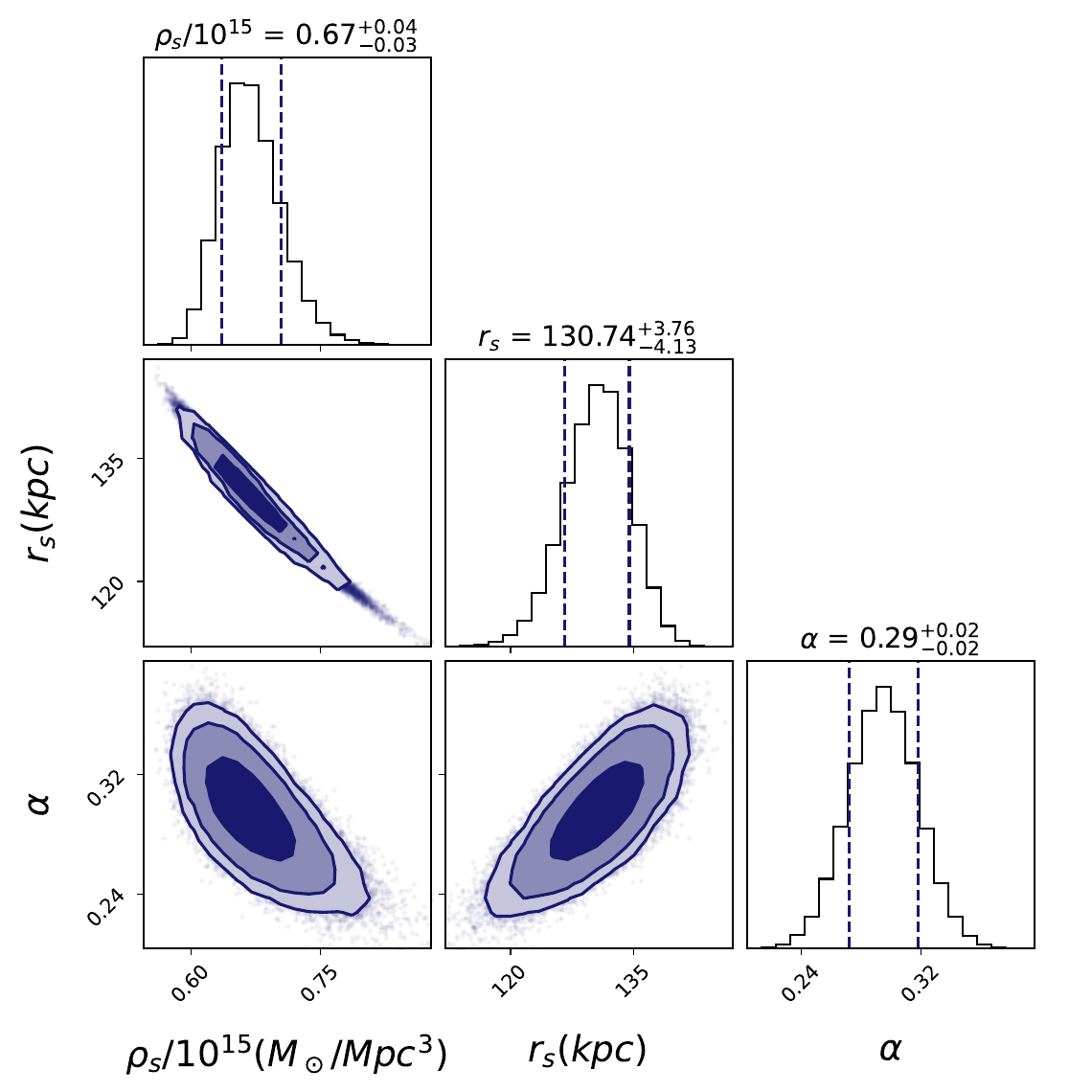}
         \subcaption{A2589}
         \label{fig:a.1}

     \end{subfigure}
     \hfill
     \hfill
     \begin{subfigure}[b]{0.48\textwidth}
         \centering
         \includegraphics[width=\textwidth]{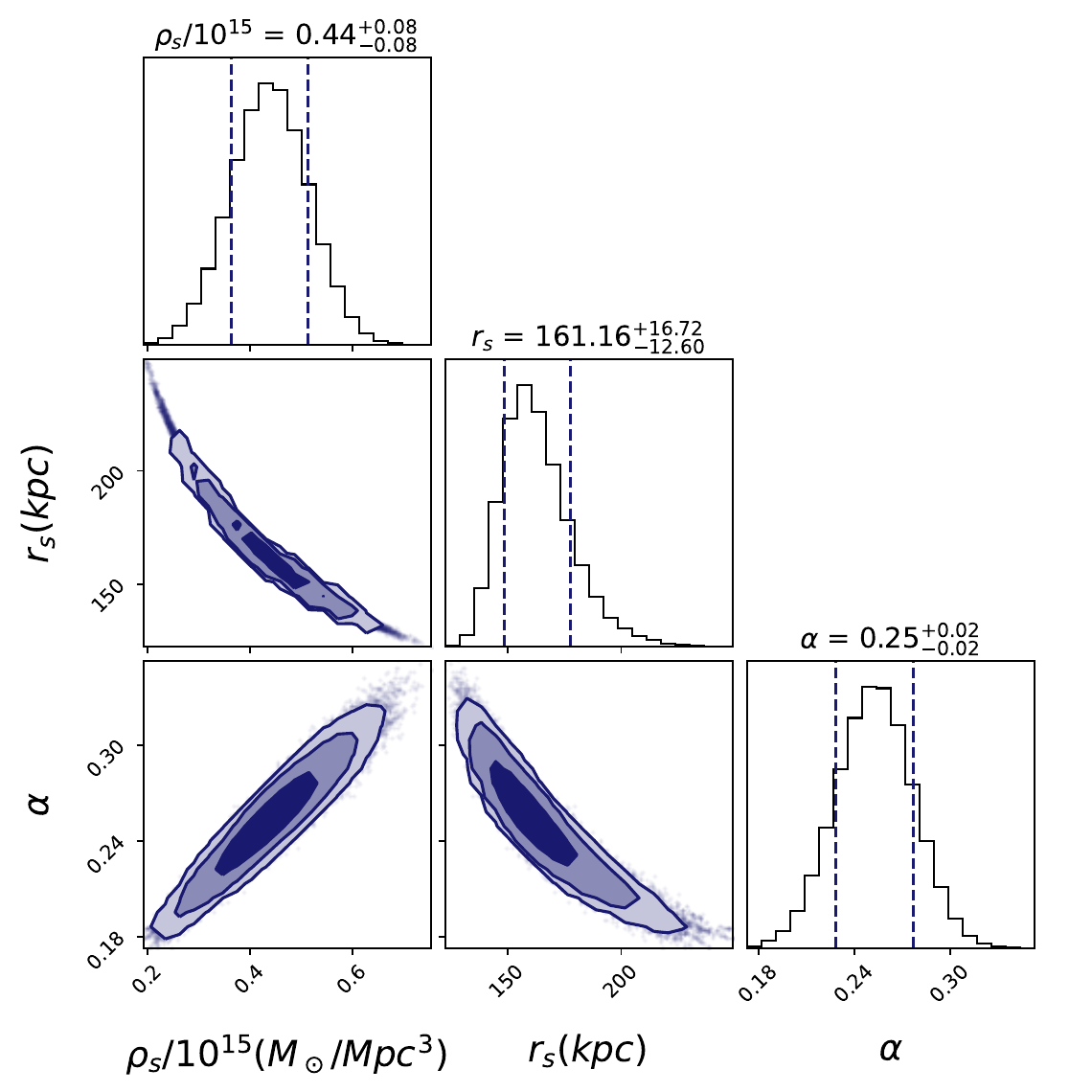}
         \subcaption{A262}
         \label{fig:a.2}
     \end{subfigure}
               \hfill
          \hfill

     \begin{subfigure}[b]{0.48\textwidth}
         \centering
         \includegraphics[width=\textwidth]{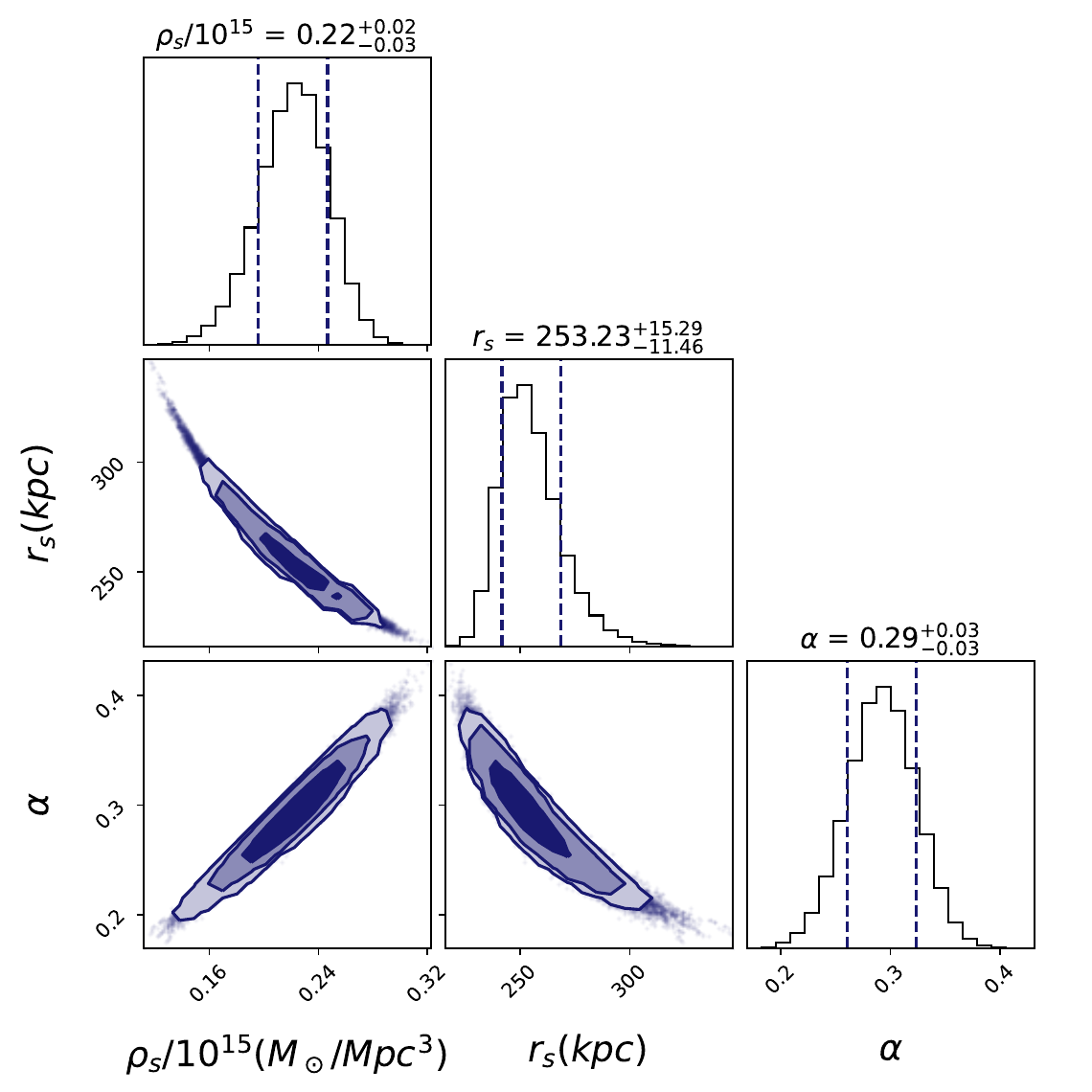}
     \subcaption{A2717}
         \label{fig:a.3}
     \end{subfigure}
     \hfill
     \hfill
     \begin{subfigure}[b]{0.48\textwidth}
         \centering
         \includegraphics[width=\textwidth]{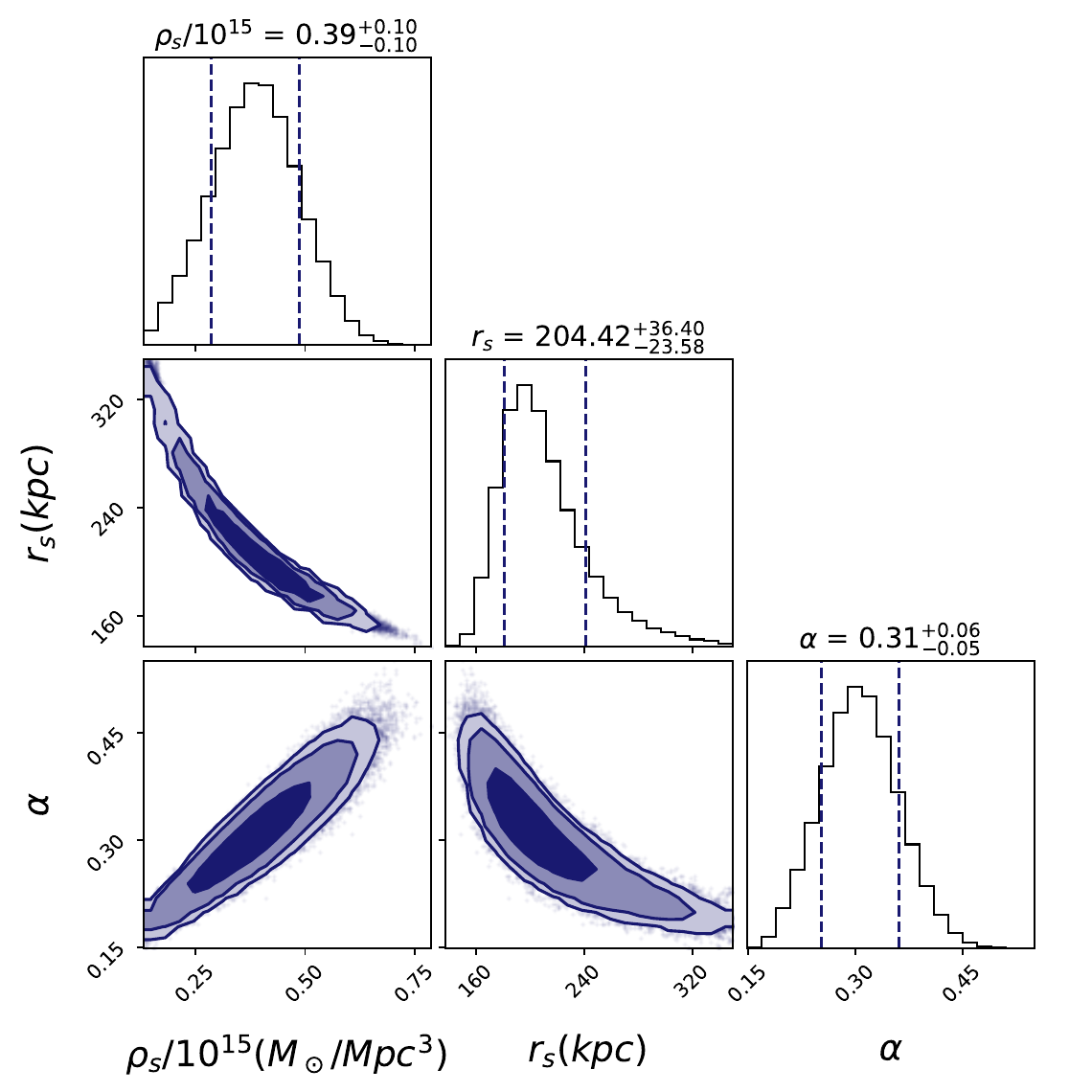}
      \subcaption{AWM 4}
         \label{fig:a.4}
     \end{subfigure}

     \hfill
     \hfill

            \caption{Corner plots for the Einasto fit to the galaxy group data. The plot shows the 68\%, 90\%, and 99\% credible intervals. Here, $\rho_s$ has  units of $M_\odot/Mpc^3$ and $r_s$ is in  $kpc$.}

        \label{fig:appendix1}
\end{figure*}

\begin{figure*}\ContinuedFloat
     \centering
     \begin{subfigure}[b]{0.48\textwidth}
         \centering
         \includegraphics[width=\textwidth]{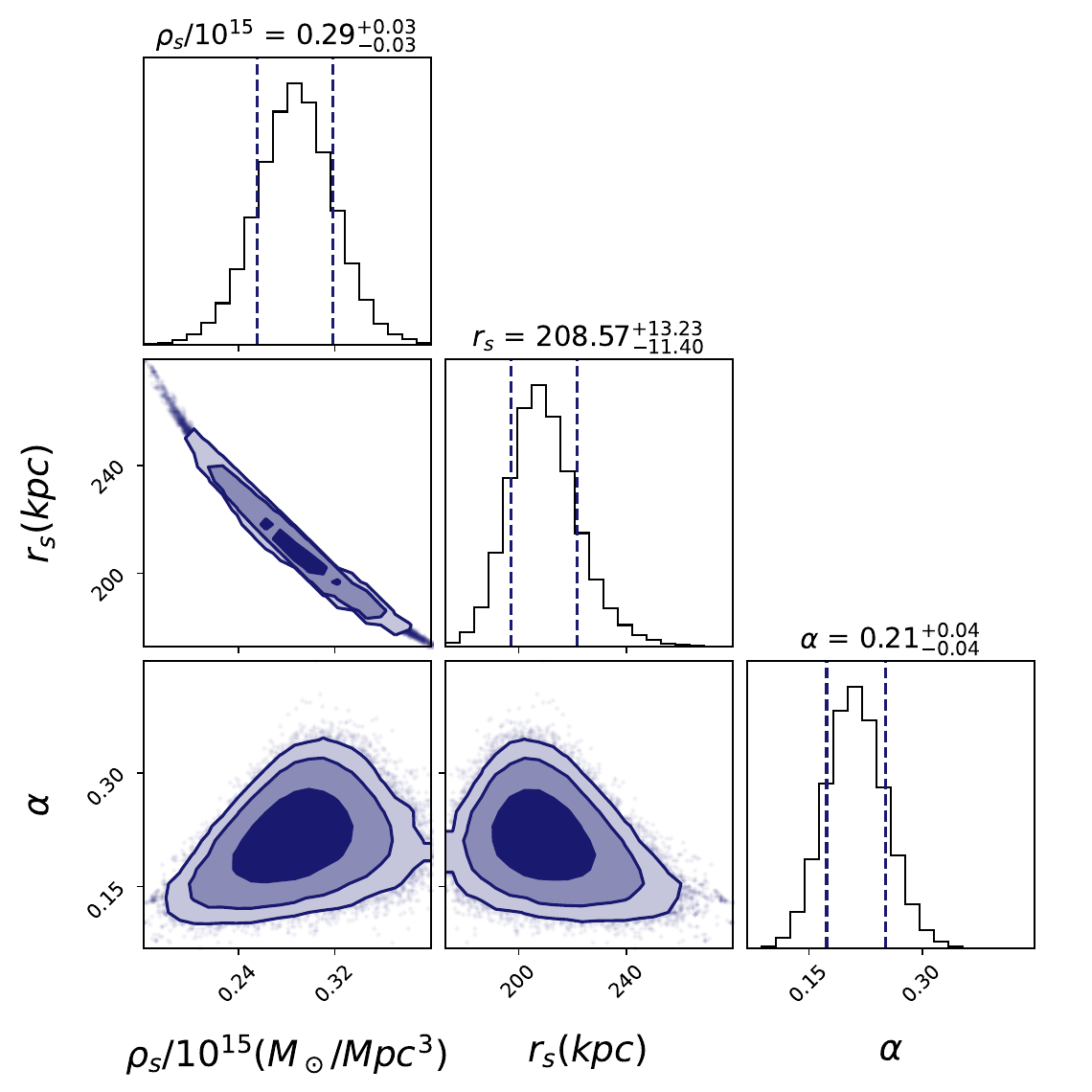}
         \subcaption{ESO 5520200}
         \label{fig:a.5}

     \end{subfigure}
    \hfill
     \hfill
     \begin{subfigure}[b]{0.48\textwidth}
         \centering
         \includegraphics[width=\textwidth]{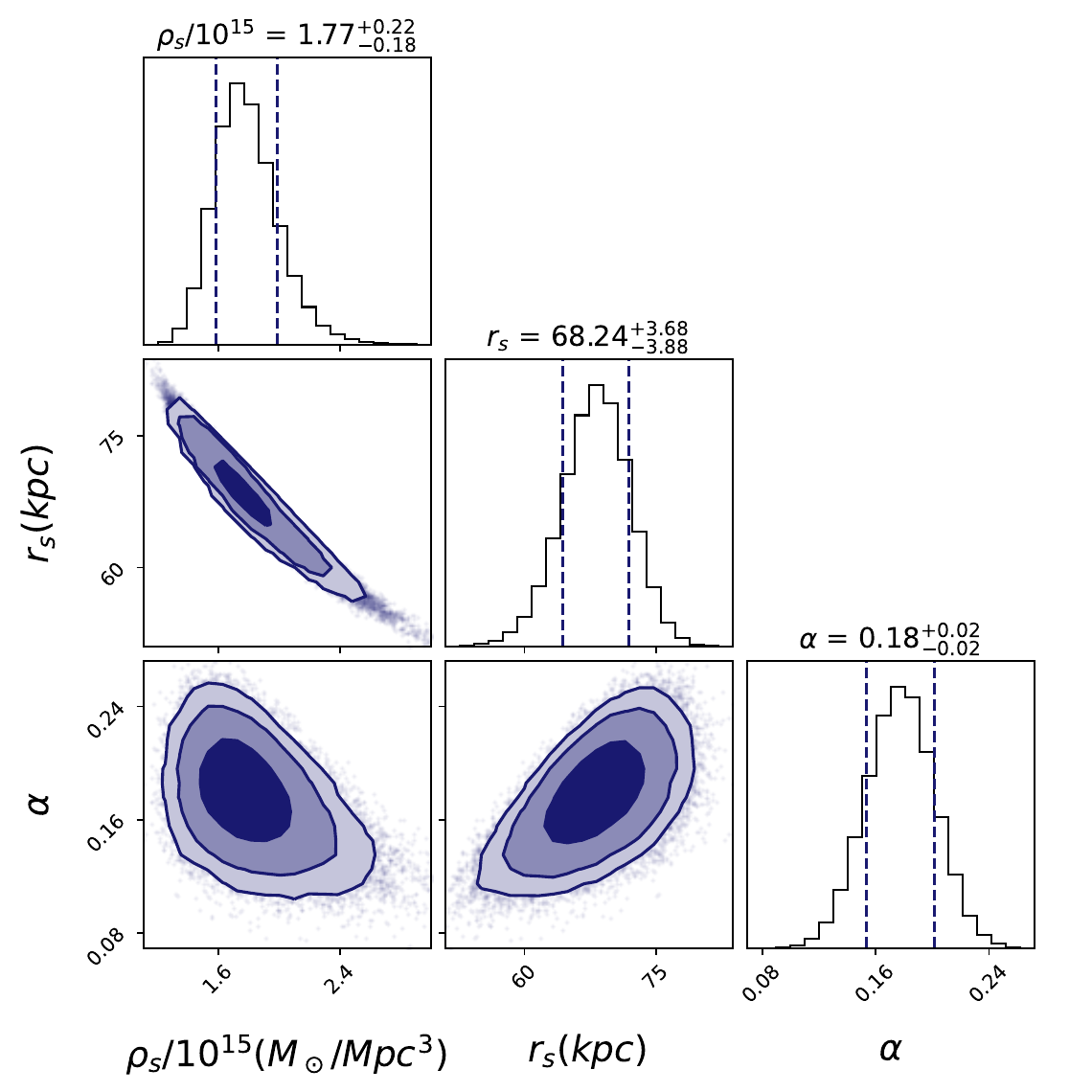}
         \subcaption{MKW 4}
         \label{fig:a.6}
     \end{subfigure}
       \hfill
     \hfill
     \begin{subfigure}[b]{0.48\textwidth}
         \centering
         \includegraphics[width=\textwidth]{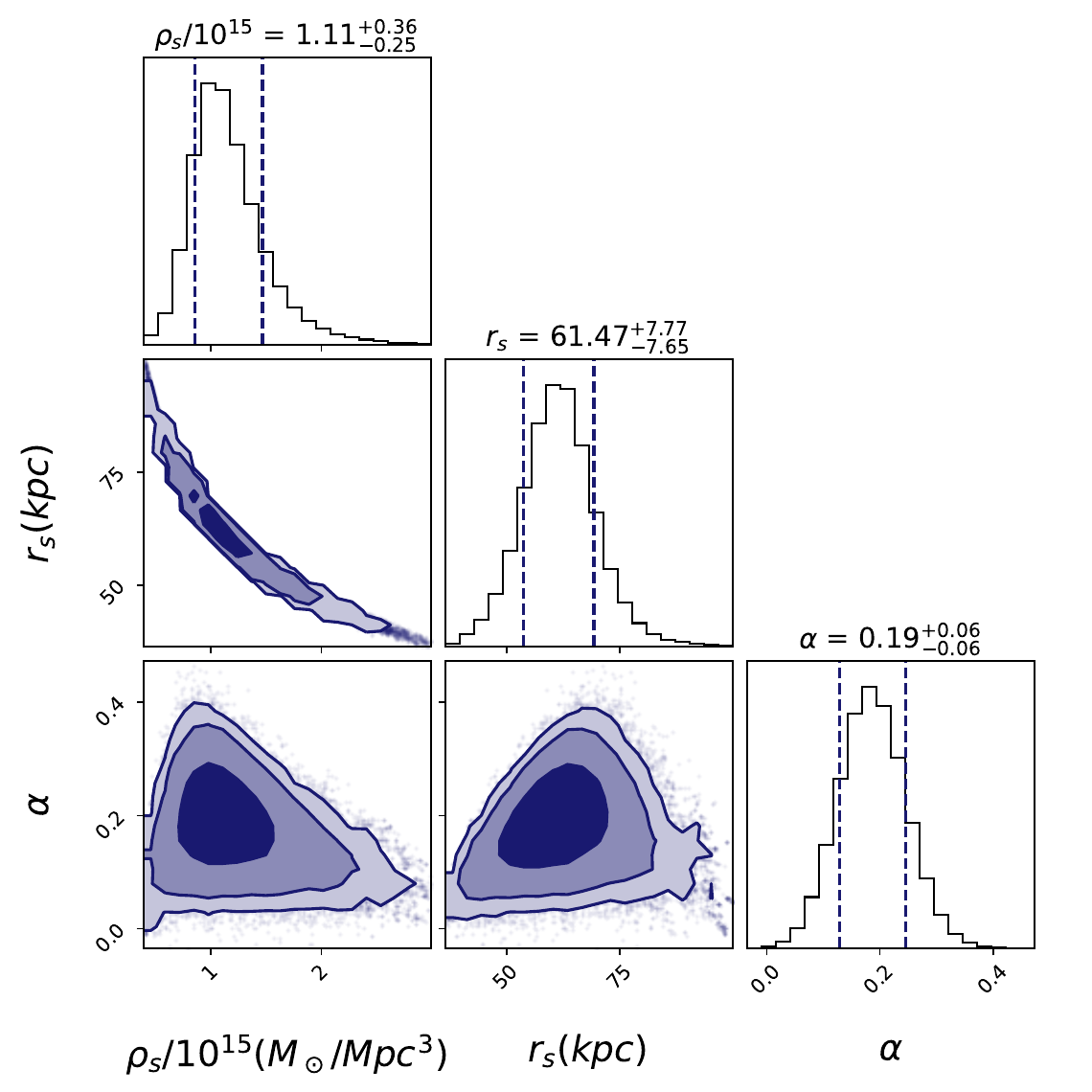}
         \subcaption{NGC 2563}
         \label{fig:a.7}
     \end{subfigure}
     \hfill
     \hfill
     \begin{subfigure}[b]{0.48\textwidth}
         \centering
         \includegraphics[width=\textwidth]{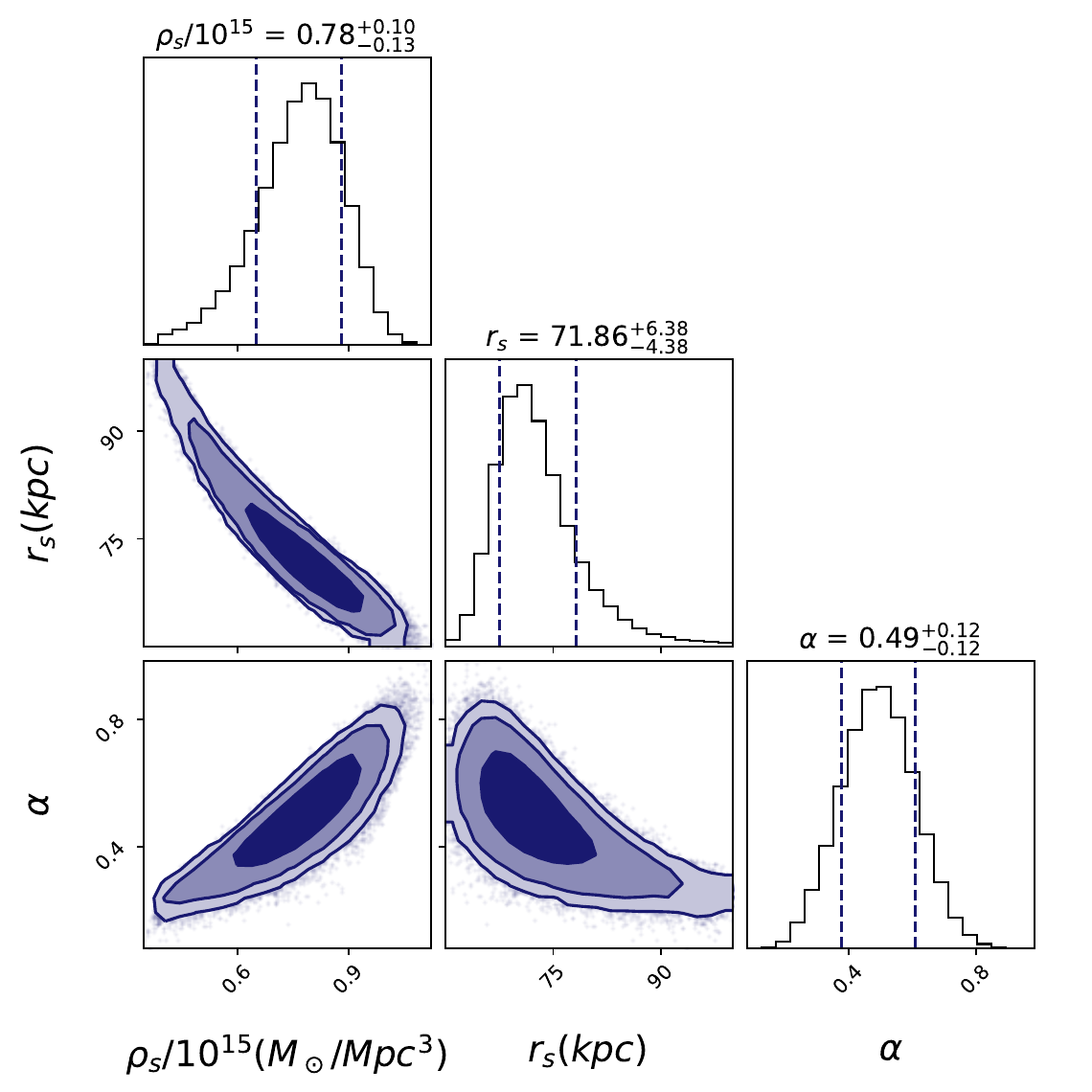}
         \subcaption{NGC 5129}
         \label{fig:a.8}
     \end{subfigure}
     \hfill
     \hfill
     
    \caption[]{Corner plots for the Einasto fit to the galaxy group data. The plot shows the  68\%, 90\%, and 99\% credible intervals. (cont.)}
    %\label{fig:appendix1}
\end{figure*}

\begin{figure*}\ContinuedFloat
     \centering
\begin{subfigure}[b]{0.48\textwidth}
         \centering
         \includegraphics[width=\textwidth]{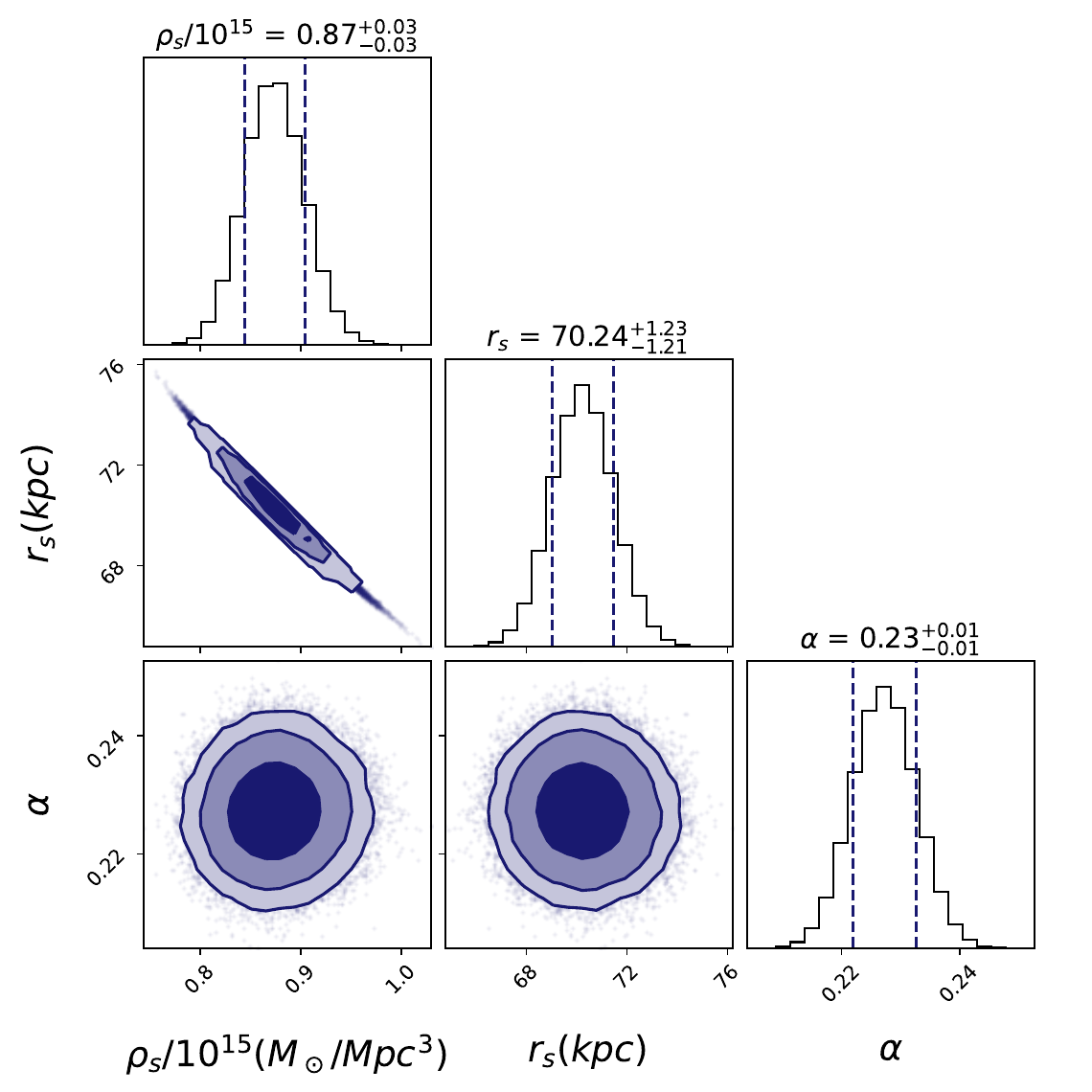}
         \subcaption{RGH 80}
         \label{fig:a.9}

     \end{subfigure}
     \hfill
     \hfill
     \begin{subfigure}[b]{0.48\textwidth}
         \centering
         \includegraphics[width=\textwidth]{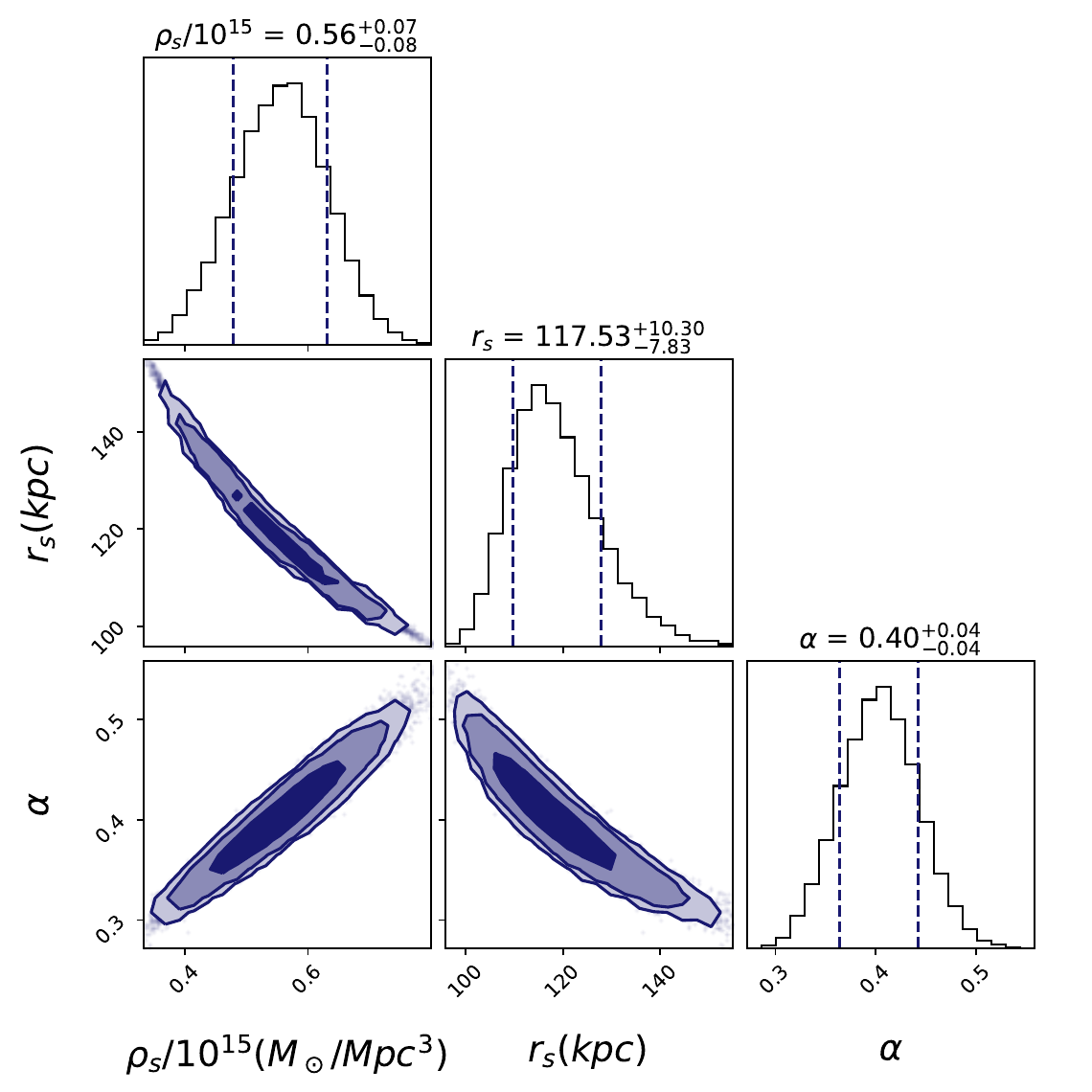}
         \subcaption{IC 1860}
         \label{fig:a.10}
     \end{subfigure}
               \hfill
          \hfill

     \hfill
     \hfill
     \begin{subfigure}[b]{0.48\textwidth}
         \centering
         \includegraphics[width=\textwidth]{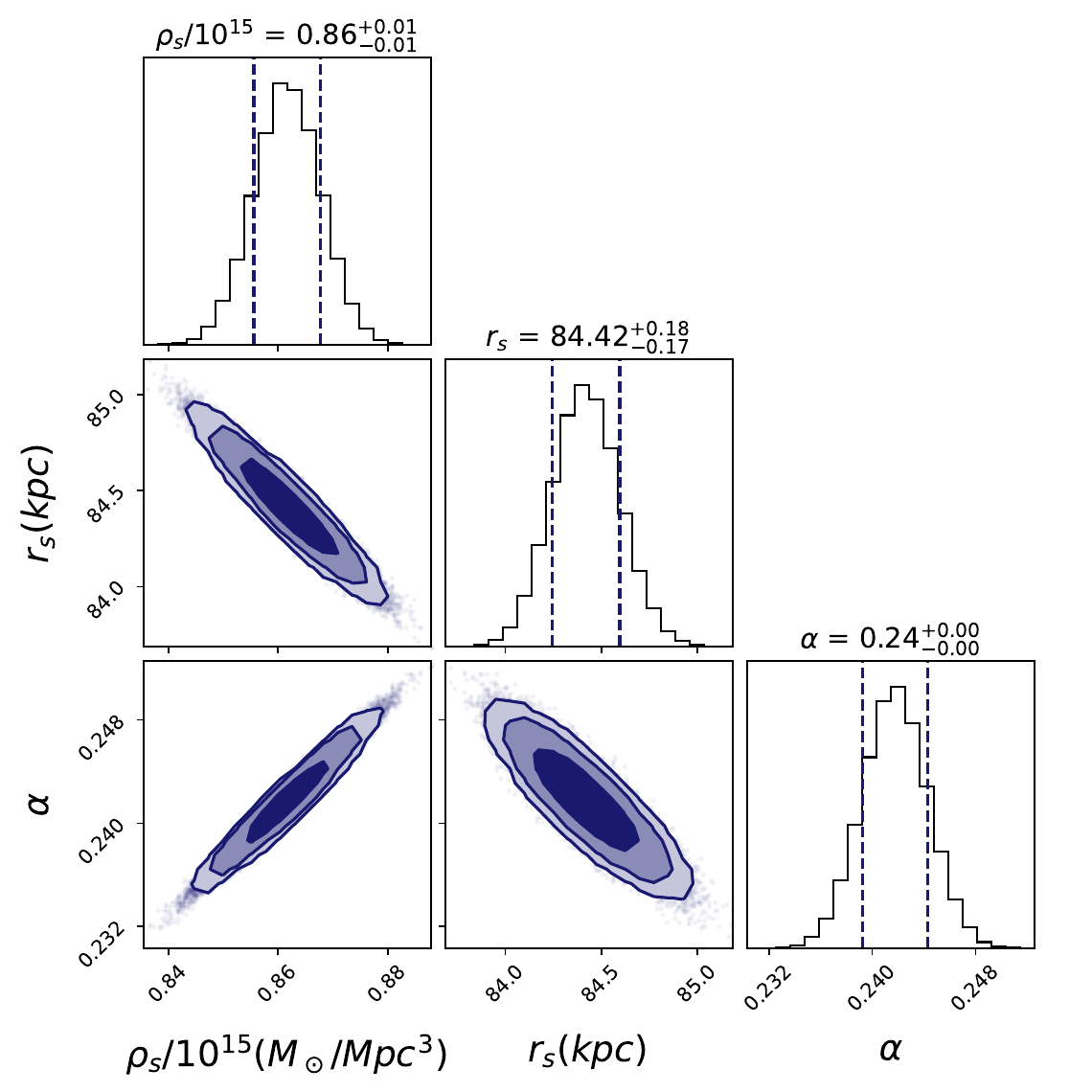}
         \subcaption{NGC 5044.  The error in $\alpha$ is 0.003. }
         
         \label{fig:a.11}
     \end{subfigure}
               \hfill
          \hfill
    
    \caption[]{Corner plots for the Einasto fit to the galaxy group data. The plot shows the  68\%, 90\%, and 99\% credible intervals. (cont.)}
    \label{fig:appendix}
\end{figure*}

\end{document}